\documentclass[aps,prc,twocolumn,superscriptaddress,showpacs]{revtex4-1}	

\usepackage{dcolumn}
\usepackage{amsmath,amssymb,amsthm,paralist}
\usepackage{graphicx}
\usepackage{appendix}
\usepackage{subfigure}
\usepackage{longtable}
\usepackage{url}
\usepackage{verbatim}

\usepackage{references}
\usepackage{hyperref} 

\usepackage{equationarray}
\usepackage[normalem]{ulem}
\usepackage[usenames]{xcolor}
\usepackage{booktabs}

\newcommand{\be}{\begin{equation}}
\newcommand{\ee}{\end{equation}}
\newcommand{\ba}{\begin{eqnarray}}
\newcommand{\ea}{\end{eqnarray}}

\begin{document}


\title{Pairing properties from random distributions of single-particle energy levels}



\author{A. A.~Mamun}
\email{ma676013@ohio.edu}
\affiliation{Department of Physics and Astronomy, Ohio University,
Athens, Ohio 45701, USA}
\author{C.~Constantinou}
\email{cc238809@ohio.edu}
\author{M.~Prakash}
\email{prakash@ohio.edu}
\affiliation{Department of Physics and Astronomy, Ohio University,
Athens, Ohio 45701, USA}


\date{\today}

\begin{abstract}




Exploiting the similarity between the bunched single-particle energy levels of nuclei and  of random distributions  around the Fermi surface, pairing properties of the latter are calculated to establish statistically-based bounds on the basic characteristics of the pairing phenomenon.  When the most probable values for the pairing gaps  germane to the BCS formalism are used to calculate thermodynamic quantities, we find that 
while  the ratio of the critical temperature $T_c$ to the zero-temperature pairing gap is close to its BCS Fermi gas value, the ratio of the superfluid to the normal phase specific heats  at $T_c$  differs significantly from its Fermi gas counterpart. The largest deviations occur when a few levels lie closely on either side of  the Fermi energy but other levels are far away from it. 
The influence of thermal fluctuations, expected to be large for systems of finite number of particles, were also investigated using a semiclassical treatment of fluctuations. When the average pairing gaps along with those differing by one standard deviations are used, the characteristic discontinuity of the specific heat at $T_c$ in the BCS formalism was transformed to a shoulder-like structure indicating the suppression of a second order phase transition as experimentally observed in nano-particles and several nuclei.   Contrasting semiclassical and quantum treatments of fluctuations for the random spacing model is currently underway.

\end{abstract}


\maketitle


\section{Introduction}

The pairing phenomenon is ubiquitous in systems of fermions interacting through  attractive interactions. 
The development of the theory for electron pairing in solids by Bardeen-Cooper-Schriffer (BCS) \cite{BCS1,BCS2} was soon followed by the realization that pairing of neutrons and protons  in nuclei led to gaps in their excitation energies \cite{BMP58}.    
Pairing also manifests itself in the binding energies of nuclei, even-even nuclei being slightly more bound than odd-even or odd-odd nuclei \cite{BM1}. Level densities of excited nuclei \cite{Bethe36,Ericson60}, their dynamical properties  such as rotational inertia \cite{Migdal59} and large amplitude motion in fissioning nuclei are also influenced by pairing \cite{RB13}. Tunneling probabilities in spontaneously fissioning nuclei are enhanced owing to pairing and thermal neutrons  induce fission of  odd-A nuclei, {\it e.g.,} $^{235}_{92}{\rm U}$ vs   $^{238}_{92}{\rm U}$.  Pairing energies in nuclei receive contributions from  sources besides BCS pairing as nuclear sizes are much smaller than the coherence length of the pairing field \cite{DH03}. The odd-even staggering is caused by a combination of effects such as the pair-wise filling of orbitals, two- and three-body interactions, the bunching of single-particle levels near the Fermi energy, and the softness of nuclei to quadrupolar interactions.  The global description of pairing in nuclei is based on the Hartree-Fock-Bogoliubov (HFB) scheme and its extensions \cite{RB13,BB05}. 
The attractive interactions between nucleons in the spin $S=0$ and $S=1$ channels are primarily responsible for pairing in nuclei.  
For accounts of recent developments in novel superfluids and superconductors in the condensed matter, nuclear and stellar environments, see Ref. \cite{NSFS}.

Through measurements of nuclear level densities $\rho\equiv\rho(E_x)$ at closely spaced excitation energies $E_x$, several attempts have been made to establish pairing correlations in nuclei  \cite{quenching,sn116,toft,pygmy}. 
The critical temperature $T_c$ at which the pairing gaps $\Delta(T)$  vanish in  systems of very large number of particles is a characteristic of a second order phase transition.  $T_c$ will be hard to pin down in nuclei as they are comprised of  small numbers of particles owing to significant fluctuations in the order parameter $\Delta(T$), but distinct signatures can likely remain.  
The experimental procedure adopted has been to examine the behavior of the specific heat at constant volume $C_V$ vs $E_x$ (or vs $T$) inferred from $\rho$ using $C_V\propto(d\ln\rho/d\ln E_x) (d\ln E_x/d\ln T)$,  
and looking for a smooth, but non-monotonic structure in $C_V$ at a critical excitation energy $E_{x,c}$ (or remnant of a critical temperature ``$T_c$'') which signals a crossover from the fully paired to the normal phase.  The moderate success achieved thus far is due to issues associated with the normalization of level densities close to the neutron separation energy \cite{Guttormsen16}. From an experimental perspective, excitation energies are well known, but not the temperature $T$ (unlike in condensed matter experiments) which requires the help of theoretical models in which the relationship between $E_x$ vs $T$ is unambiguous, albeit model dependent.  (Hereafter, we will drop the quotes in ``$T_c$'' for simplicity, but it should be understood as referring to the 
temperature around which a non-monotonic structure in $C_V$ vs $T$ is present.)  Additional complications arise for $T\leq T_c$ due to the role of collective effects which influence the magnitude of $\rho$.  Notwithstanding these difficulties, the goal of establishing $T_c$ in nuclei appears to be within reach through continuing innovations in experimental techniques and  theoretical efforts. Indeed, a shoulder-like structure (also referred to as an S-shape structure, although a severe bending of one's head is required to see the S in many cases)  in $C_V$ vs $T$ has been experimentally observed for many nuclei  \cite{quenching,sn116,toft,pygmy}.

The study of fluctuations in the order parameter $\Delta(T)$ and the suppression of superconductivity/superfluidity in systems of  small number of particles (nano-particles in modern parlance) in condensed matter physics \cite{Anderson59,Falci00}   predates similar efforts in nuclear physics \cite{Morettoplb,Al13}. The inadequacy of the mean-field BCS formalism becomes apparent in situations when the mean level spacing of the single-particle (sp) energy levels $\delta \gtrsim \Delta$.  As these studies have revealed, the absence of a second order phase transition with a discontinuity in $C_V$ at $T_c$ is direct consequence  of large fluctuations in $\Delta$. A study of the role of thermal fluctuations, albeit  with a semiclassical treatment of fluctuations following Refs. \cite{LLI,Morettoplb}, is also undertaken in this work. This treatment goes beyond BCS insofar as the gauge (number) symmetry broken in the BCS approach is restored. A full quantum treatment of fluctuations is outside the scope of this work, but will be reported separately.

In this work, we introduce the random spacing (RS) model to study the pairing properties of a system consisting of a finite number of nucleons.  The basic feature of the RS model is the randomly distributed  (sp) energy levels around the Fermi surface to mimic the bunched shell-model orbitals in nuclei  generated through the use of different underlying energy density functionals.  
Although reminiscent of the random matrix model, the RS model differs from it in that diagonalization of a random Hamiltonian matrix is bypassed.  Insofar as many random realizations of the sp energy levels will be considered for a fixed number of particles, the use of different physically motivated energy density functionals leading to different disposition of the sp levels will be captured. 
The pairing properties of the RS model are explored in two distinct  stages as outlined below.

In the first stage, the BCS formalism in which the most probable gap values are employed to calculate thermodynamic quantities such as the excitation energy, entropy and specific heat is used. 
The ensuing results are compared with the analytical results of the Fermi gas (FG) and constant spacing (CS)  models as well those of select nuclei. In the second stage, the role of fluctuations is examined based on a semiclassical treatment of fluctuations reserving for a later study a fully quantum treatment of the same.  As in the first stage, a comparison with results of the CS model and those of nuclei including fluctuations is made.

The organization of this paper is  as follows. In Sec.~\ref{RSmodel}, the basic features of the RS model are introduced.  A description of the theoretical approach in the first stage of our investigations and a discussion of our results is contained in Sec.~\ref{Form1}. The influence of thermal fluctuations on the pairing gap and on the thermodynamic quantities  examined in the second stage is described in Sec.~\ref{Form2}, which also includes results and discussion. Our summary and conclusions are in Sec.~\ref{Sumconc}.

\section{The Random Spacing Model}
\label{RSmodel}

\begin{figure}[!htb]
\centerline{\includegraphics[width=1.0\columnwidth,angle=0]{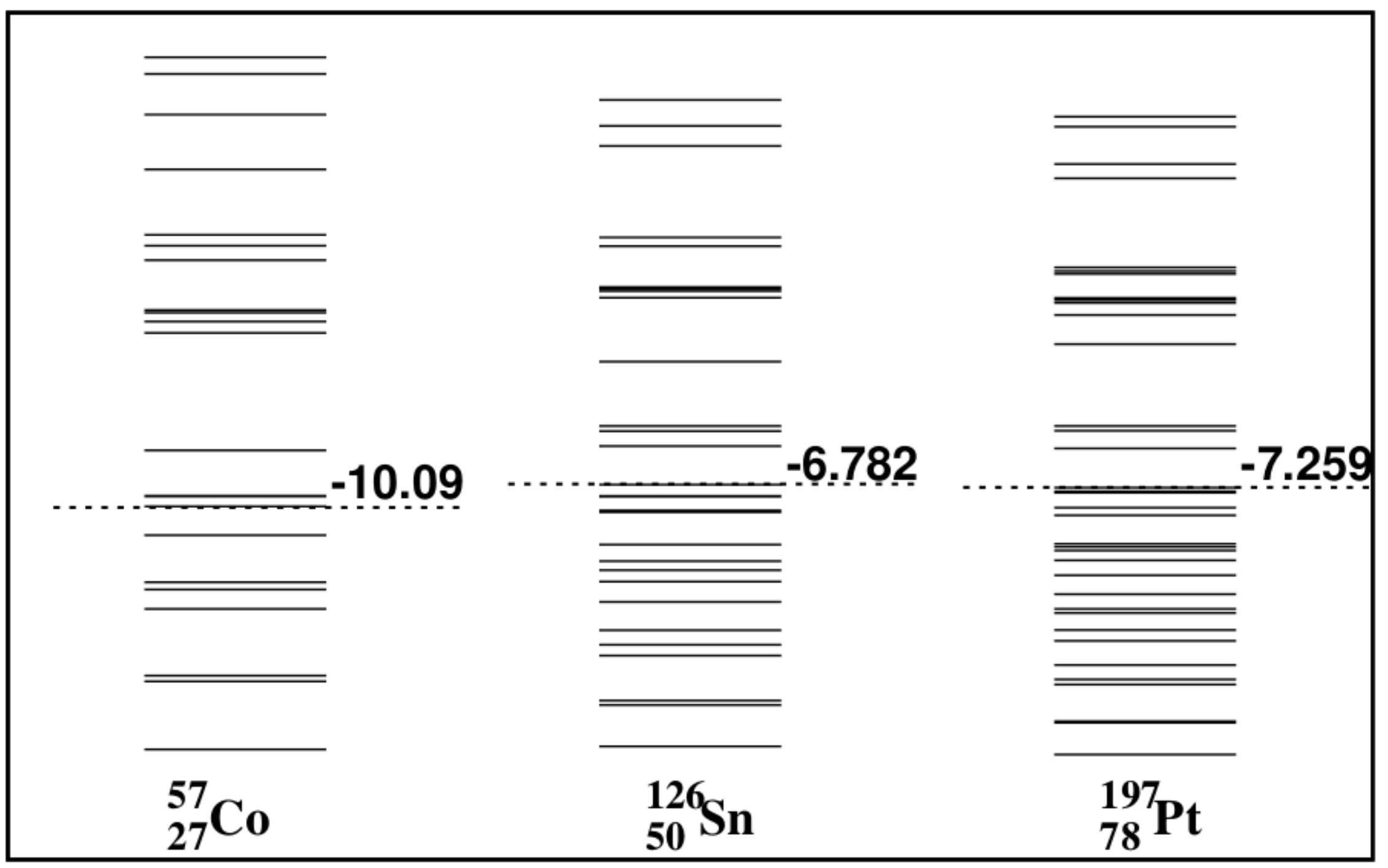}}
\vskip -0.25cm
\caption{Neutron single-particle energy levels in the indicated nuclei
from  HFB calculations \cite{Reinhardbook,Reinhardpaper} using the SkO$^\prime$ energy density functional with full pairing. The dotted lines indicate the location of the Fermi energies in each case.}
\label{nsplevels}
\end{figure}

Our objective here is to examine the pairing properties in a global manner keeping in mind that the single-particle (sp) energies of nuclei 
exhibit bunching caused by shell and pairing effects.  
Figure \ref{nsplevels} shows the bunching of neutron sp energies  from  HFB calculations 
using the energy density functional SkO$^\prime$ 
with full pairing in  $^{57}$Co, $^{126}$Sn and $^{197}$Pt \cite{Reinhardbook,Reinhardpaper}. The proton levels for these cases (not shown) also exhibit  similar bunching.  We stress, however, that use of different energy density functionals and pairing schemes (constant force, surface or bulk pairing, etc.) lead to 
significant differences in the spacing of levels around the Fermi surface \cite{reinhardShape}.

\begin{figure}[!htb]
\centerline{\includegraphics[width=1.0\columnwidth,angle=0]{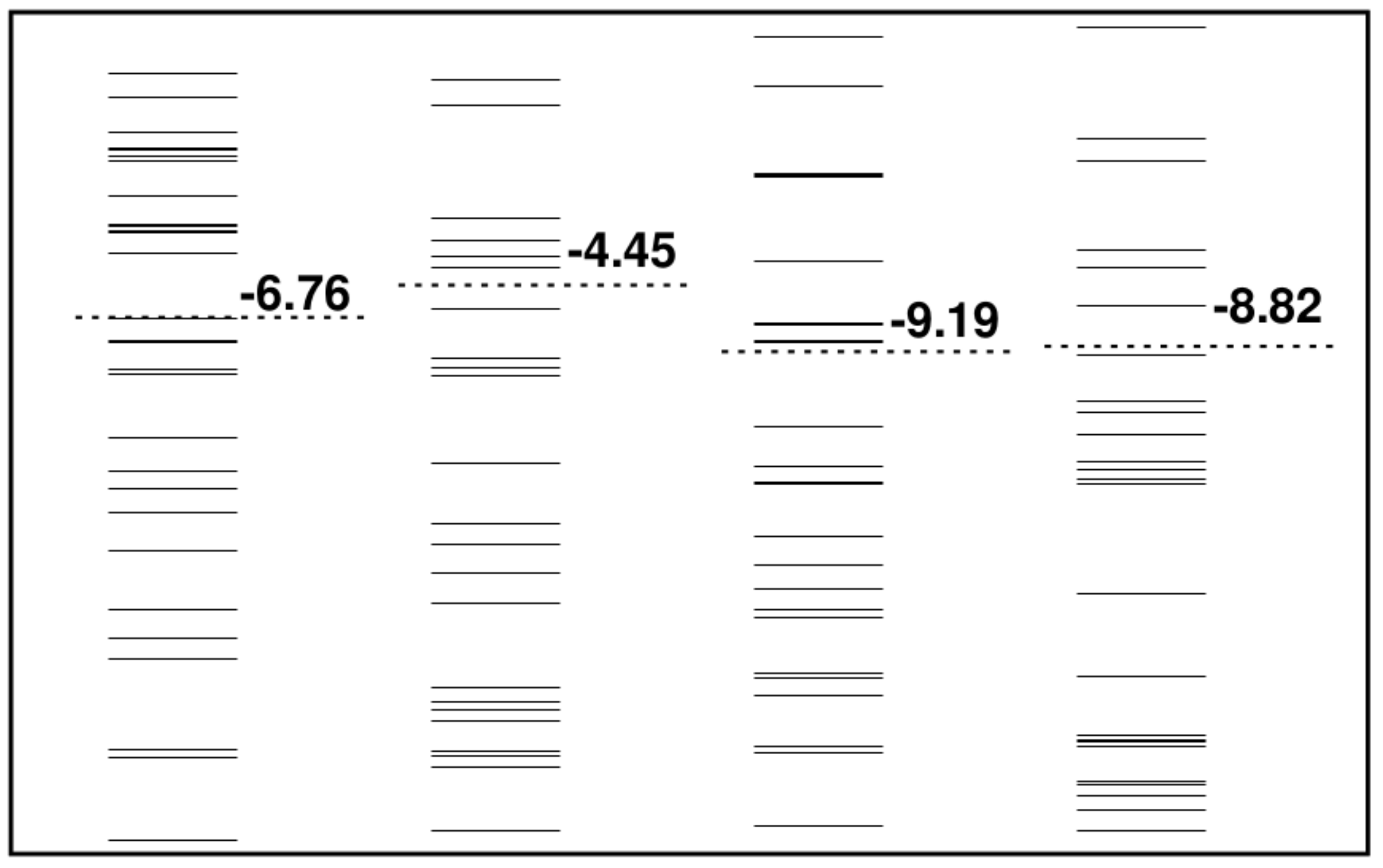}}
\vskip -0.25cm
\caption{Neutron single-particle energy levels in  $\rm ^{126}Sn$  from HF+BCS calculations  \cite{Reinhardbook,Reinhardpaper} using the SkO$^\prime$ energy density functional with constant pairing force (leftmost set) and three randomly
generated single-particle energy levels.
\label{random}}
\end{figure}

When the sp levels of a large number of nuclei are examined, they appear to resemble 
those generated randomly around the Fermi surface. 
An example is shown in Fig. \ref{random} where the neutron sp levels of 
$\rm ^{126}Sn$ are contrasted with three cases of randomly generated sp levels with the 
same number of neutrons at $T=0$.  Although not exact replicas, the latter share the property
of bunched levels with nuclei.  In a set consisting of a very large number of randomly generated sp levels for a given nucleus, 
some are likely to represent the true situation, especially considering the dependence on different energy density functionals currently in use.  
Thus, the primary focus of this work is to examine the pairing properties from randomly distributed  
sp energy levels with appropriate constraints imposed to model  sp energy levels of nuclei.  
We will 
(i) address the extent to which the basic characteristics such as $ {T_c}/{\Delta_0}$ (where $\Delta_0=\Delta(T=0)$), the ratio of superfluid to normal specific heats at constant volume, $\left. {C_V^{(s)}}/{C_V^{(n)}}\right|_{T_c},  $ and $ \frac {1}{T_c} \left. \frac {d\Delta^2}{dT}\right|_{T_c}$ compare with those  Fermi gas (FG) and HFB calculations, and  
(ii) place statistically-based bounds for the case randomly distributed sp energy levels.

\section{Pairing Properties}
\label{Form1}

With model sp energies as input, various physical quantities can be calculated utilizing the 
BCS equations generalized to include angular momentum \cite{Morettonucl}:
\begin{eqnarray}
\label{numeqn}
 N &=& \sum_{s,k}   \left[ 1 - \frac{\epsilon_{k} - \lambda}{2E_{k}} \tanh \left(\frac{E_{k}+(-1)^s\gamma m_k}{2T}\right) \right]  \\
   \label{gapeqn}
  \frac {2}{G} &=& \sum_{s,k} \frac {1}{2E_k} \tanh \left(\frac{E_{k}+(-1)^s\gamma m_k}{2T}\right) \,, \\
  \label{Meqn}
  M &=& \sum_{s,k} m_k \frac {(-1)^{s+1}}{1+\exp \left(\frac{E_{k}+(-1)^s\gamma m_k}{2T}\right)   } \,,
\end{eqnarray}
where  $N$ denotes the number of particles, $\epsilon_k$ are the sp energies, 
$\lambda$ is the chemical potential, and $T$ is the temperature. The summation index  $s$ takes on the values 1 and 2, whereas the index $k$  sums over all sp energy levels. The quasi-particle energy is 
\begin{equation}
E_k = \sqrt{(\epsilon_k - \lambda)^2 + \Delta^2} \,,
\end{equation}
where $\Delta$ is the  pairing gap at the Fermi surface generated by the pairing interaction with strength $G$. 
The quantity $M$ is the projection of the total angular momentum on a laboratory-fixed $z$-axis or on a body-fixed $z^\prime$-axis, $m_k$ are the sp spin projections and $\gamma$ is the Lagrange multiplier that fixes $M$.  
Equations (\ref{numeqn})-(\ref{Meqn}), studied  as a function of $(T,M)$  for fixed $(N,G)$ provide the critical temperature $T_c$ below which the system is paired $[\Delta\equiv \Delta(T,M) \neq 0]$ and above which it is normal $[\Delta = 0]$. The excitation energy $E_x=E(T)-E(0)$, entropy $S$ and the specific heat at constant volume $C_V=T(dS/dT)$ are obtained from \cite{Morettonucl,SY63}
\ba 
\label{BCSEx}
 E(T) &=& \sum_{s,k} \epsilon_{k} \left[ 1 - \frac{\epsilon_{k} - \lambda}{E_{k}} \tanh \left(\frac{E_{k}^{(s)}}{2T}\right) \right] 
 - \frac{\Delta^{2}}{G}\\
\label{BCSentropy}
 S &=&  \sum_{s,k} \left\{\ln [1 + \exp (-E_{k}^{(s)}/T)] + \frac{E_{k}^{(s)}/T}{1 + \exp(E_{k}^{(s)}/T)}\right\} \nonumber \\ \\
 C_V &=& \frac{1}{4}\sum_{s,k} \frac{E_k^{(s)}/T}{{\rm cosh}^2(E_{k}^{(s)}/2T)} 
  \left[\frac{E_{k}^{(s)}}{T} - \frac{1}{2E_k} \frac{d\Delta^2}{dT} \right]  \,,
\label{BCSCV}
\ea
where $E_k^{(s)} = E_k + (-1)^{s} \gamma m_k$. 

To mimic the sp energy levels $\epsilon_k$ of nuclei in the random spacing (RS) model, 
random numbers from a uniform sequence are generated between $\pm 2\hbar\omega$ from the Fermi energy $E_F (=\lambda~\rm {at}~T=0$) with 
$\hbar\omega = 41A^{-1/3}$, where $A$ is the mass number, to conform to 
the systematics of spacing between major shells in nuclei \cite{BM1}. 
For light nuclei, Ref. \cite{Blom67} recommends the relation $\hbar\omega = 45A^{-1/3} - 25A^{-2/3}$.
HFB and/or HF+BCS calculations of nuclei guide the choice of 
$G$  in solving Eqs. (\ref{numeqn}) and (\ref{gapeqn}) 
to obtain $\Delta$ and $\lambda$.  In the results reported below, the $T=0$ pairing energies were tallied with the systematics for nuclei with $A=N+Z$ \cite{GFB13}. For the neutron pairing gaps,
\ba
\Delta_{N,Z} &=& 24/A + 0.82 \pm 0.27~{\rm MeV}, \quad {\rm for}~ N~{\rm odd}, \nonumber \\ 
\Delta_{N,Z} &=& 41/A + 0.94 \pm 0.31~{\rm MeV}, \quad {\rm for}~ N~{\rm even}\,, 
\label{nsystematics}
\ea
whereas for the proton pairing gaps,
\ba
\Delta_{N,Z} &=& 0.96 \pm 0.28~{\rm MeV}, \quad {\rm for}~ Z~{\rm odd}, \nonumber \\ 
\Delta_{N,Z} &=& 1.64 \pm 0.46~{\rm MeV}, \quad {\rm for}~ Z~{\rm even}\,. 
\label{psystematics}
\ea

As in the case of the FG or constant spacing (CS) models, an analytical calculation of  $Q=-\frac {1}{T_c}\left. \frac {d\Delta^2}{dT}\right|_{T_c}$  \cite{BCS2,SY63} is precluded for discrete bunched levels. 
We have therefore devised a 3-term formula utilizing Refs. \cite{AS,RG27}  
to calculate $Q$.  Explicitly, for a general $f$ and step size $h$, the right end-point derivative is 
\ba
 f^\prime(x) &=& \frac {1}{39} \left[ 32 \phi \left(\frac h4 \right) + 12 \phi \left(\frac h2 \right) - 5 \phi \left(h\right) \right]  + {\cal O}(h^4) \,,
 \nonumber \\
 \phi(h) &=& \frac {1}{2h} \left[ f(x-2h) - 4 f(x-h) + 3 f(x) \right] \,.
\ea

\subsection*{Results}
\label{Res1}

First, we recall  the analytical results for spin-doublet  sp levels (degeneracy $d=2$) in
the FG model in which the sp level density $g(\epsilon) \propto \epsilon^{1/2}$  and
for the CS model in which $g(\epsilon)$ is a constant, for both of which \cite{BCS1,BCS2,Morettonucl,SY63,ll9}
\begin{eqnarray}
\Delta_0 &=& \frac {\hbar \omega}{\sinh (1/gG)} \approx 2 \hbar \omega \exp\left(-\frac {1}{gG} \right) \\
\frac {T_c}{\Delta_0} &\simeq& 0.57, ~~ \frac {C_V^{(s)}}{C_V^{(n)}} \simeq 2.43 ~~ {\rm and} ~~ - \frac {1}{T_c} 
\left. \frac {d\Delta^2}{dT}\right|_{T_c} \simeq 9.4 \,, \nonumber \\
\label{FGprops}
\end{eqnarray}
where $\pm\hbar\omega$ are the upper and lower limits of integration above and below the Fermi energy $E_F$.  
The similarity of results in these two models stems from 
the conditions  $T/E_F<<1$, $\Delta_0/E_F << 1$ (i.e., pairing  
is a Fermi surface phenomenon) and $g(E_F)G<<1$ (weak coupling) being satisfied.

\begin{figure}[!htb]
\begin{center}
\includegraphics[width=1.0\columnwidth,angle=0]{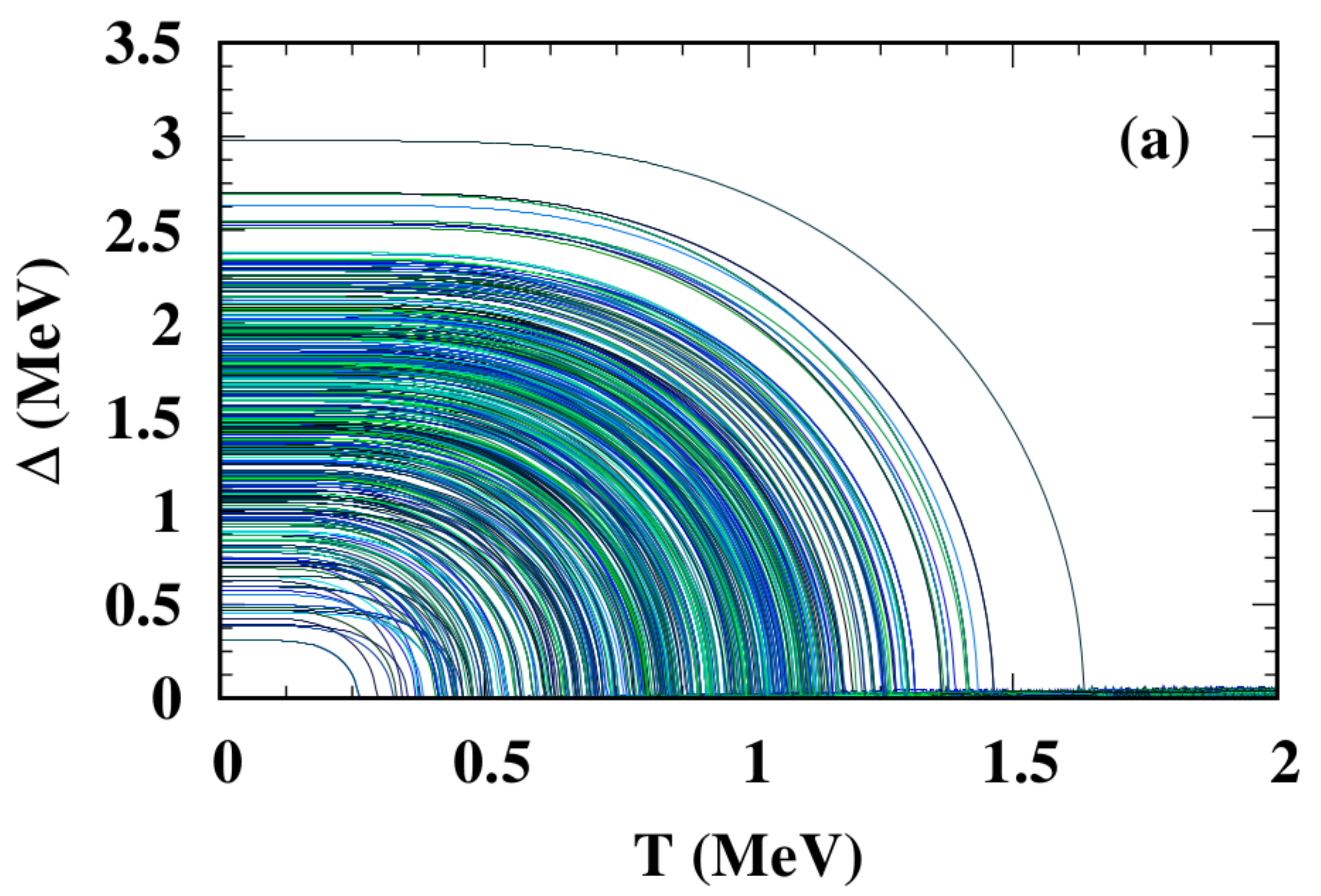} 
\vskip -0.5cm
\includegraphics[width=1.0\columnwidth,angle=0]{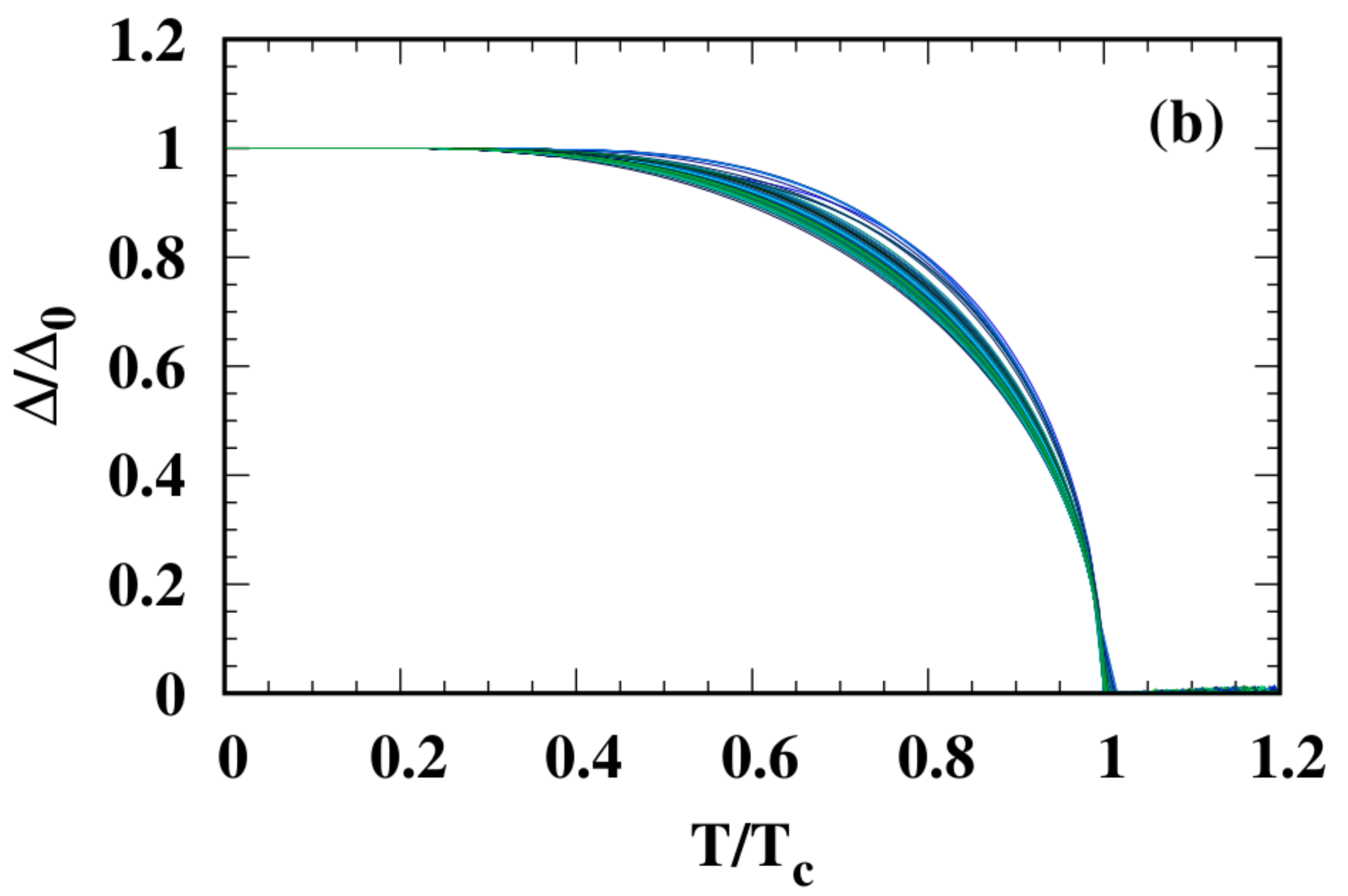} 
\end{center}
\vspace{-0.5cm}
\caption{(Color online.) (a) Pairing gap vs temperature for 500 sets of  randomly 
generated sp levels.
(b) Pairing gap normalized to its zero-temperature value vs temperature normalized to 
the critical temperature for the results in (a).}
\label{deltaAll}
\end{figure}

Our results for the RS model are for a large number ($\geq 500$) of independent random realizations of sp energy levels for a given $N$ at $T=0$.  
The pairing gap $\Delta$ vs  $T$  shown in Fig. \ref{deltaAll}(a) corresponds to 500 
such realizations with  $N=76$,  
$M=0$  and $G=0.2$ (similar results ensue for other values of $G$) for spin-doublet levels. 
The varying $\Delta_0$ and $T_c$ are due to 
the different set of sp levels encountered in each run. Every curve in this figure 
resembles the BCS prediction for the FG or CS  model. The nearly universal behavior  
 displayed in Fig. \ref{deltaAll}(b) indicates
that even for randomly generated sp levels, 
deviations from the BCS relations 
\ba
\Delta/\Delta_0&\simeq& 1 - \left(\frac{2\pi T}{\Delta_0}\right)^{1/2}\exp\left(-\frac{\Delta_0}{T}\right) \quad {\rm for}~~ T<<\Delta_0\,, \nonumber \\
 &\simeq& 1.74\left(1-\frac{T}{T_c}\right)^{1/2} \quad {\rm for}~~ T_c-T<<T_c \,,
\ea
are small. Small quantitative differences  from the BCS result for  intermediate values of $T/T_c$,
evident from the band-like structure of the 
bell-shaped curve in  Fig. \ref{deltaAll}(b), are caused by the variety of levels close to $E_F$.   

%

\begin{figure}[!htb]
\begin{center}
\includegraphics[width=1.0\columnwidth,angle=0]{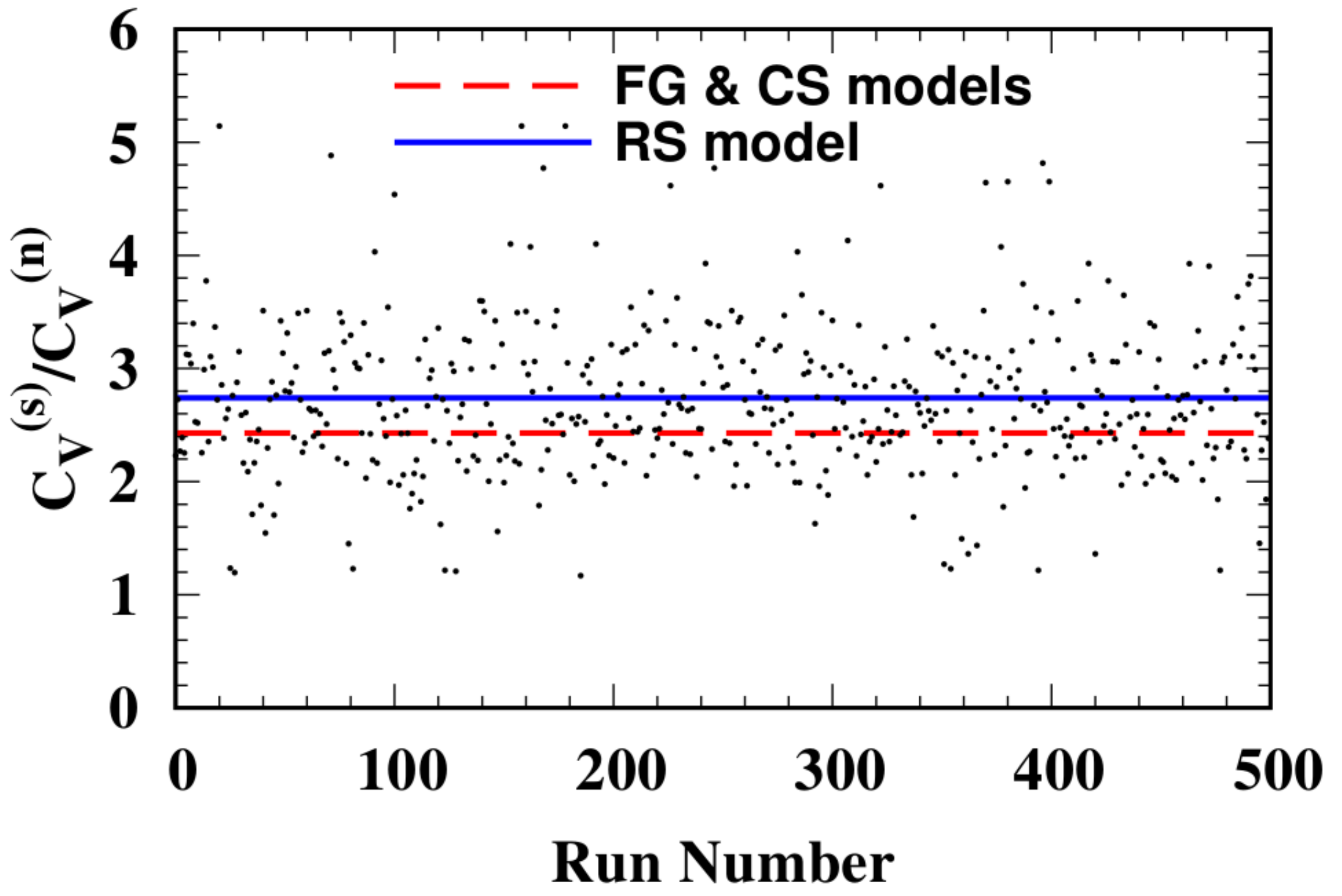}
\vspace{-0.5cm}
\caption{(Color online.) Ratio of the superfluid to normal phase specific heats at constant volume at $T_c$.
\label{cvratio}}
\end{center}
\end{figure}

The ratio $ {C_V^{(s)}}/{C_V^{(n)}}$ at $T_c$
is shown in Fig. \ref{cvratio}. Note that the scatter around the mean value, which is 
moderately close to that for the FG or CS model, is significant for  the RS model. 
The outlying points in this figure correspond to cases in which a couple of  levels lie closely on either side of $E_F$, but other levels are far away from it.  
In Table \ref{table1}, the basic characteristics of the phase 
transition for the RS model are compared with those of FG and CS  models.

\begin{figure}[!htb]
\begin{center}
\includegraphics[width=1.0\columnwidth,angle=0]{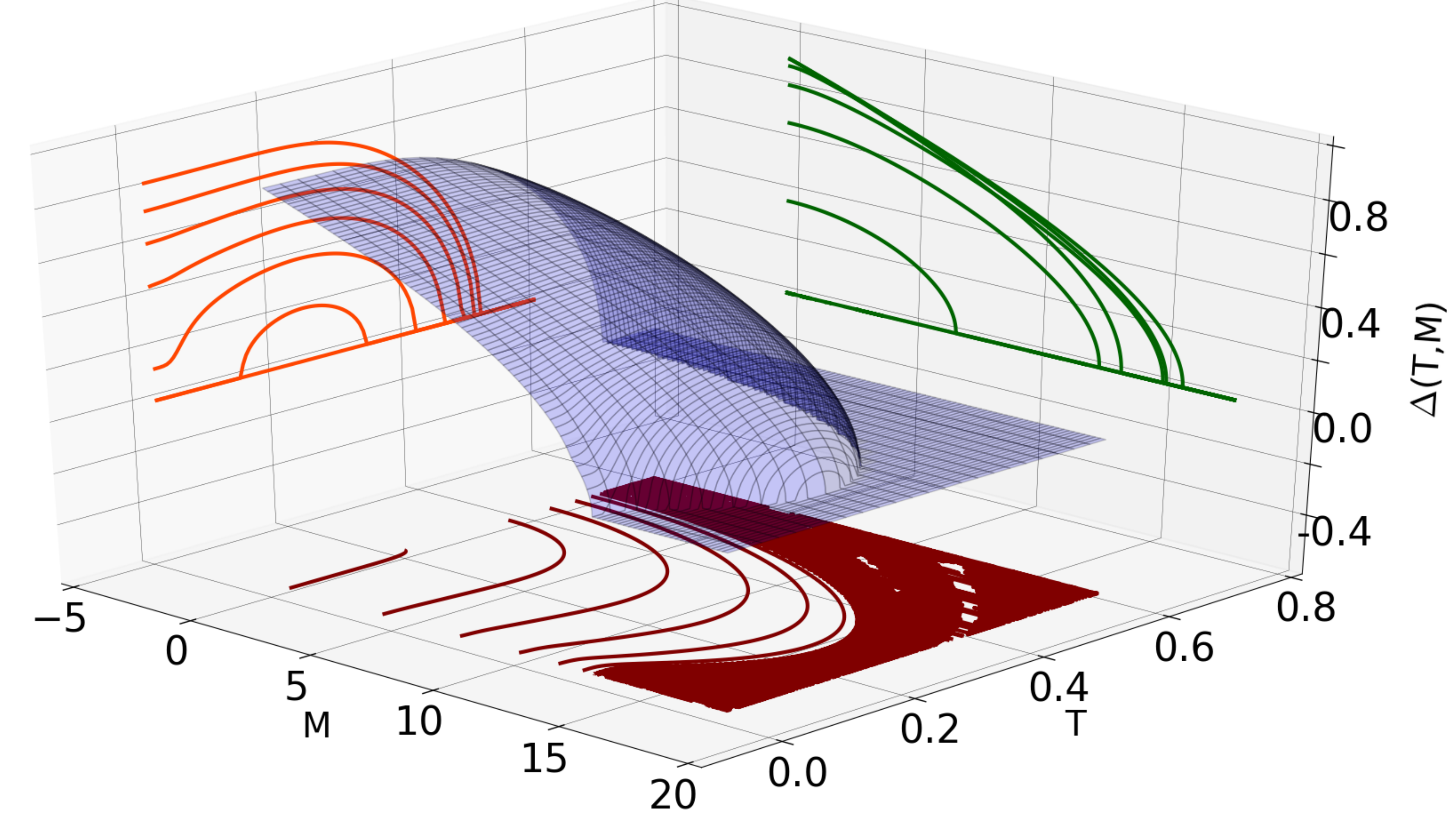}
\vspace{-0.5cm}
\caption{(Color online.) Pairing gap $\Delta$ vs temperature, $T$, and  the projection of the total angular momentum, $M$. 
\label{3d}}
\end{center}
\end{figure}

\begin{table}[!htb] 
\caption{Characteristics of the  pairing phase transition. Results for the CS and RS models with $N=76$ are for 1000 runs. The HF+BCS results for the SkO$^\prime$ energy density functional are for protons and neutrons, respectively. Entries with N/A correspond to the case when $\Delta_0=0$.}
\begin{center} 
\begin{tabular}{lrrr}
\hline
\hline
Model & $\frac{ T_c}{\Delta_0}$ & $\left.\frac{C_V^{(s)}}{ C_V^{(n)}}\right|_{T_c}$ 
&Q= -$\frac{1}{T_c}\left.\frac{d\Delta^2}{dT}\right|_{T_c}$\\
\hline
FG \& CS($d=2$)& $\simeq 0.57$  & $\simeq 2.43$  & $\simeq  9.4$ \\
CS($d=2j+1$) & $0.55$  & $2.99$  & $9.92$ \\
RS($d=2$) & $0.57 \pm 0.05$  & $2.71 \pm 0.73$  & $9.51 \pm 0.81$ \\
RS($d=2j+1$) & $0.57 \pm 0.04$  & $3.05 \pm 1.53$  & $9.55 \pm 0.98$ \\
\hline 
HF+BCS($^{57}$Co) & 0.59,0.56 & 2.30,2.33 & 9.85,9.75 \\
HF+BCS($^{126}$Sn) & N/A,0.54 & N/A,4.47 & N/A,10.03 \\
HF+BCS($^{197}$Pt) & 0.55,0.54 & 3.46,4.93 & 10.33,10.52 \\
\hline
\hline
\end{tabular}
\end{center}
\label{table1}
\end{table}


The role of angular momentum on $\Delta$ is shown in Fig. \ref{3d} for the RS model with $\Delta_0=1$ MeV and $m_k=2$ to provide comparison with similar results for the CS model \cite{Morettonucl}.   Increasing values of $T$ and $M$ diminish $\Delta$, and thus $T_c$ relative to when $M=0$.
As for the CS model, the paired region extends to $M_{max}$ beyond $M_c$ at which $T_c=0$ in the RS model with 
$M_{max}/M_c =  1.22 \pm 0.12$ in accord with $\simeq1.22$ for the former case.  
For $M_c<M<M_{max}$, two critical points exist in both of these models. For values of $M$ accessible in  experiments, this region is likely not encountered. We have verified that up to $M\simeq10$, the first order approximation, $M \simeq (\beta \gamma /2) \sum_k
m_k^2~{\rm sech}^2(E_k/2T)$, is sufficiently accurate, terms involving higher even powers of $m_k$ being required only for larger $M$.  

\begin{figure}[!htb]
\begin{center}
\includegraphics[width=1.0\columnwidth,angle=0]{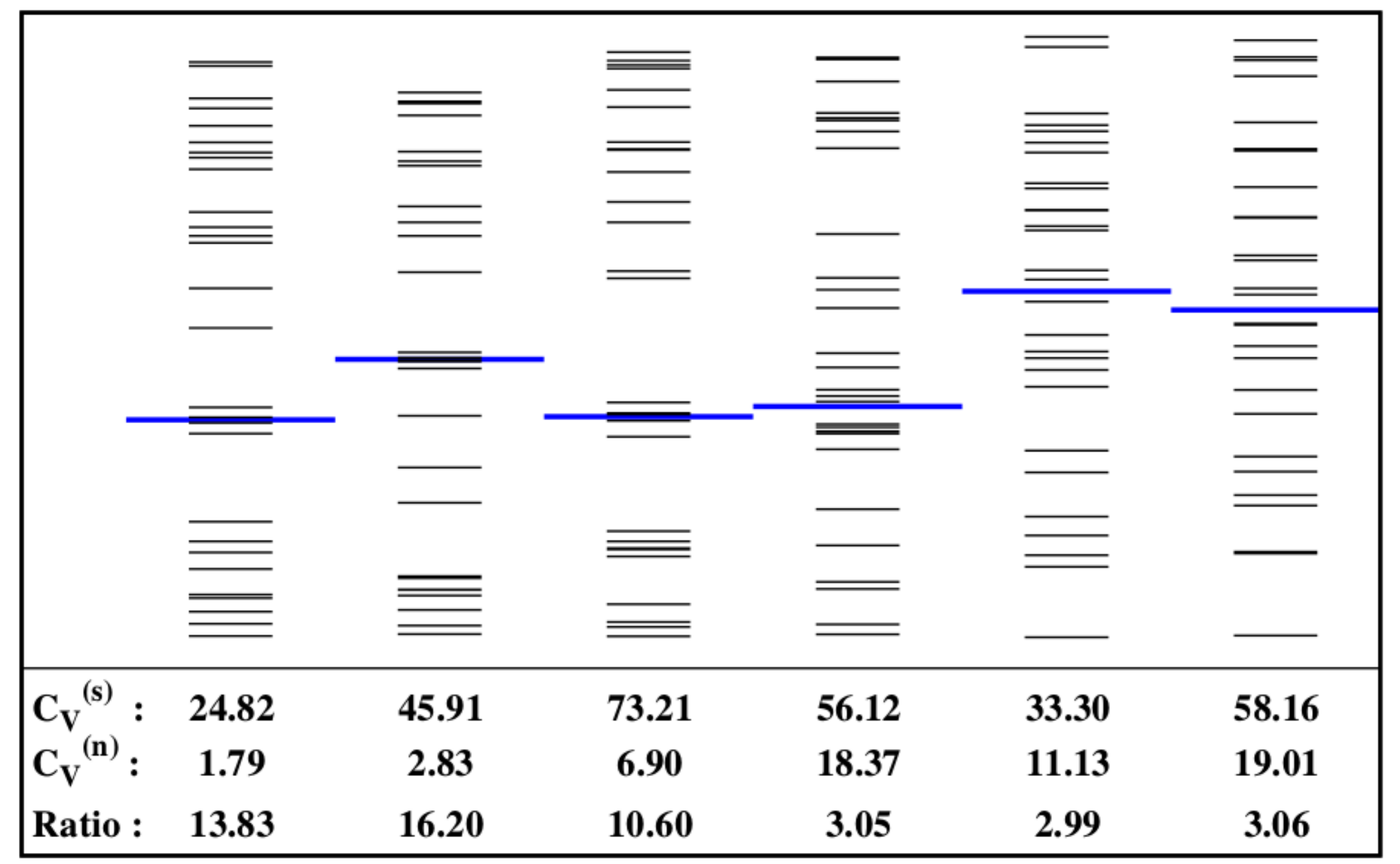}
\vspace{-0.5cm}
\caption{(Color online.) Examples of random sp energy spectra illustrating the origin of large deviations in 
$C_V^{s}/C_V^{n}$. The long horizontal lines show the locations of the corresponding Fermi energies. 
\label{devs}}
\end{center}
\end{figure}

Endowing the sp energy levels of the CS and RS models with  $d=2j+1$ angular momentum $(j)$ degeneracies of the shell model orbitals of spherical nuclei yield results similar (to within the standard deviations shown in Table \ref{table1}) to those for nuclei.
The $T=0$ results for the nuclei shown  conform to the nuclear systematics in Eq. (\ref{nsystematics}). 
Results of the RS model for values of $N$ other than 76 show similar trends. For example, $T_c/\Delta_0 = 0.56\pm0.04~(0.57 \pm 0.04),~C_V^{(s)}/C_V^{(n)}=3.23\pm 1.47~(3.46\pm2.15)$, and $Q=9.81\pm 1.07~(9.71\pm1.17)$ for $N=30~(119)$.  
Note that although $T_c/\Delta_0$'s remain close to the FG or CS model predictions, properties associated with the specific heat vary considerably in the RS model as well as in HF+BCS calculations owing to the variety of bunched sp levels encountered.  
The largest deviations from the mean values occur when a few levels are on either side of $\lambda$, but other levels are far away from it (see Fig.~\ref{devs}). In such cases, $C_V^{(n)}$ is significantly smaller than those in other cases which renders the ratio $C_V^{(s)}/C_V^{(n)}$ very large. 

Unlike in the FG and CS models in which the parameter $gG$ chiefly determines the pairing properties,  the average density of states $\bar g(E_F)$ and $G$ separately influence results  in the RS model as well as those in  HF+BCS calculations of nuclei.  By comparing with results of  HFB calculations for typical cases,   
we have verified that results of the RS model encompass the case of deformed nuclei for which the above degeneracies are lifted.

\section{Influence of Fluctuations in $\Delta$}
\label{Form2}

The gap equation, Eq. (\ref{gapeqn}), follows from the condition 
\begin{equation}
\left. \frac {\partial \Omega}{\partial \Delta}\right|_T = 0\,,
\label{gapcond}
\end{equation}
where (we take $M=0$ hereafter for simplicity) the function
\begin{eqnarray}
\Omega(T,\Delta) &=& -  \sum_k \frac{(\epsilon_k - \lambda - E_k)}{T} \nonumber \\
&+&  2 \sum_k \ln \left[1 + \exp\left(- \frac{E_k}{T}\right) \right]  - \frac{\Delta^2}{GT}
\label{omega}
\end{eqnarray}
determines the grand partition function ${\cal Z}=\exp(\Omega)$. Equation (\ref{gapcond}) delivers the most probable gap values $\Delta_{\rm mp}(T)$. The transition to the paired state is usually a second order phase transition and $\Delta_{\rm mp}(T)$ is its order parameter.  Its decrease with increasing $T$ is continuous with a discontinuity in its slope at the critical temperature $T_c$ beyond which the system becomes unpaired. There is no latent heat but a discontinuity in specific heat at $T_c$.   Utilizing $\Delta_{\rm mp}$ to determine the thermal variables is justified when the corresponding probability distribution $P(\Delta)$ is sharply peaked at $\Delta_{mp}$. In a system with a large number of particles, $P(\Delta)$ approaches a delta function. 
However, nuclei are comprised of a small number of particles and fluctuations can be very large, particularly when the mean single-particle  level spacing $\bar \delta = 1/\bar g \gtrsim \Delta$. In this case, superconductivity/superfluidity is expected to vanish although pairing correlations may persist.  For small number of particles, but still with $\Delta \sim \delta$, quantum fluctuations suppress superconducting properties and the mean field BCS theory becomes invalid.
These features were uncovered for small superconducting grains (nano particles) in condensed matter physics~\cite{Anderson59,Falci00} and are also characteristic of nuclei with small number of particles~\cite{Al13}. 

When $\Delta \gg \delta$ and can be considered as strongly coupled to all the other intrinsic degrees of freedom, the isothermal semiclassical probability distribution for $\Delta$ is given by \cite{LLI,Morettoplb}  
\begin{equation}
P(\Delta) \propto \exp[ \Omega(T,\Delta) ] \,.
\label{PDelta}
\end{equation}

As emphasized in Ref. \cite{LLI}, when the temperature is too low or when $\Delta$ varies too rapidly with time the fluctuations cannot be treated thermodynamically, and a quantum treatment becomes necessary to account for the purely quantum fluctuations.   
Here we will use the semiclassical treatment of fluctuations in $\Delta$ as in Ref. \cite{Morettoplb} by  using Eq. (\ref{PDelta}) to examine the extent of its utility and also to identify the regions of $T$ or $E_x$ in which a proper quantum treatment is necessary.  

In what follows, we consider the role of fluctuations in $\Delta$ on the thermal variables for the CS, RS and HFB models when the gap values 
\begin{eqnarray}
\Delta_{\rm av} &=& \frac {\sum_\Delta \Delta~ P(\Delta)}{\sum_\Delta P(\Delta)}, \quad \Delta_{\rm av} \pm \sigma \quad {\rm with} 
\nonumber \\
\sigma &=&\left[ \frac {\sum_\Delta \Delta^2~ P(\Delta)}{\sum_\Delta P(\Delta)} - \Delta_{\rm av}^2 \right]^{1/2}
\end{eqnarray}
are used to calculate the various thermal variables. 
For any value of $\Delta$ including $\Delta_{mp}$, the number, energy, and entropy expressions (for $M=0$) ) are given by \cite{Morettonucl,Morettoplb}
\begin{eqnarray}
N = \frac {\partial \Omega}{\partial \alpha} &=&
\sum_k  \left[ 1 - \frac{\epsilon_{k} - \lambda}{E_{k}} \tanh \left(\frac{E_{k}}{2T}\right) \right] \nonumber \\
 &+& \frac {\Delta} {T} \frac {\partial \Delta}{\partial \alpha} 
 \left[ \sum_k \frac {1}{E_k} \tanh \left(\frac{E_k}{2T} \right]  - \frac 2G  \right) \,,
 \label{Nfluc}
\end{eqnarray}
where $\alpha=\lambda/T$,  
\begin{eqnarray}
 E &=& T^2 \frac {\partial \Omega}{\partial T} = \sum_k \epsilon_{k} \left[ 1 - \frac{\epsilon_{k} - \lambda}{E_{k}} \tanh \left(\frac{E_{k}}{2T}\right) \right] 
 - \frac{\Delta^{2}}{G} \nonumber \\
 &-& \left(\Delta^2  - \Delta T \frac {\partial \Delta}{\partial T} \right) 
 \left[ \sum_k \frac {1}{E_k} \tanh \left(\frac{E_k}{2T} \right)  - \frac 2G  \right] 
 \label{Efluc}
\end{eqnarray}
and 
\begin{eqnarray}
S&=&\Omega + (E-\lambda N)/T \nonumber \\
 &=&2 \sum_k \left\{\ln [1 + \exp (-E_k/T)] + 2~ \frac{E_k/T}{1 + \exp(E_k/T)} \right\} \nonumber \\
 &-& \frac {\Delta}{T} \left( \frac {\lambda}{T} \frac {\partial \Delta}{\partial \alpha} - T \frac {\partial \Delta}{\partial T} \right)  
 \left[ \sum_k \frac {1}{E_k} \tanh \left(\frac{E_k}{2T} \right)  - \frac 2G  \right] \,.  \nonumber \\
 \label{Sfluc}
\end{eqnarray}
For $\Delta=\Delta_{\rm mp}$, the familiar forms for these quantities are recovered as the factor in the last parenthesis in each of the above expressions is the gap equation in Eq. (\ref{gapeqn}) for $M=0$  which vanishes.  The specific heat at constant volume  $C_V=dE/dT=T(\partial S/\partial T)$ is readily evaluated numerically (or from the lengthy analytical expression in \cite{Kargar13}).
The numerical results presented below for $E_x$ and $C_V$ are for $E_x(\Delta_{mp}, \Delta_{av}, \Delta_{av} \pm \sigma)$ and  $C_V(\Delta_{mp}, \Delta_{av}, \Delta_{av} \pm \sigma)$, respectively, where appropriate. Note, however, that
\begin{equation}
\langle Q \rangle = \frac {\sum_\Delta Q~ P(\Delta)}{\sum_\Delta P(\Delta)} \neq Q(\Delta_{av}) \,,
\end{equation}
except when $Q$, that can be any of $E_x, C_V$ and $S$, is  a linear function of $\Delta$ which is not true in the present context.  
Nonetheless, the results shown below amply illustrate the role of fluctuations in $\Delta$.  

Note that the last terms in Eqs. (\ref{Nfluc}) through (\ref{Sfluc}) involving the gap equation together with appropriate multiplicative factors takes the semiclassical analysis of fluctuations beyond BCS, but remains at the mean field level insofar as only thermal fluctuations on a static underlying mean field are considered.  Equation (\ref{Nfluc}) ensures number conservation thereby restoring the broken gauge (number) symmetry of the BCS approach. 

\subsection*{Results}
\label{Res2}

\subsection*{The CS Model}

\begin{figure}[!htb]
\begin{center}
\includegraphics[width=1.0\columnwidth,angle=0]{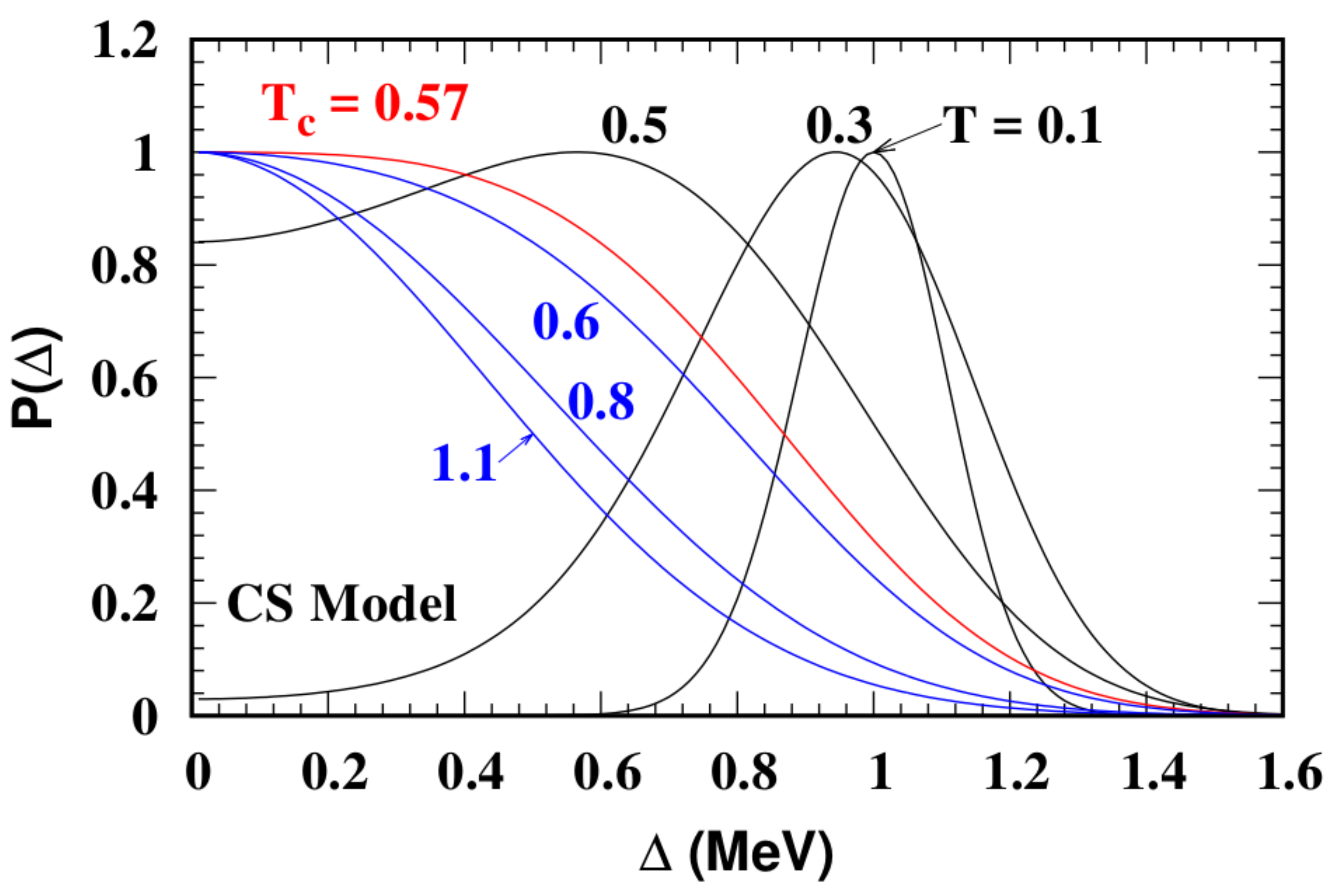}
\vspace{-0.5cm}
\caption{(Color online.) Probability distributions $P(\Delta)$ vs $\Delta$ at different $T$'s. The maximum for each $P(\Delta)$ occurs at the corresponding $\Delta_{{\rm mp}}$ obtained from Eq. (\ref{gapeqn}).
\label{CSprob}}
\end{center}
\end{figure}

Although the influence of fluctuations in the CS model have been considered before using the semiclassical treatment described above in Ref. \cite{Morettoplb}, we summarize our main findings here to enable comparison with the results in the RS and HFB models to be discussed later. We also include results related with standard deviations from $\Delta_{\rm av}$ not shown in Ref. \cite{Morettoplb}.    
 The role of fluctuations is analyzed by choosing a constant spacing $g=5~{\rm MeV}^{-1}$ between doubly degenerate single-particle levels for $A=144$ and $\Delta_0=1~{\rm MeV}$  at $T=0$ as in Ref. \cite{Morettoplb}. For this choice, $G=0.0581$ MeV, $\hbar \omega\simeq 41 A^{-1/3}=7.78$ MeV,  with levels distributed between $\pm 2\hbar\omega$ around  $\lambda_{\rm mp}(0)=-1.3471$ MeV at $T=0$.  Figure \ref{CSprob} shows $P(\Delta)$  (normalized such that $P(\Delta_{{\rm mp}})=1$)  vs $\Delta$ for different temperatures.   Noteworthy features in this figure are: (i) For $T\simeq 0$, the distribution $P(\Delta)$  is symmetrical around $\Delta_{{\rm mp}}$, (ii) with increasing $T$, $P(\Delta)$ becomes increasingly asymmetrical, and (iii) For $T\geq T_c\simeq 0.57$ MeV, $P(\Delta)$ is peaked at $\Delta=0$.  
 
\begin{figure}[!htb]
\begin{center}
\includegraphics[width=1.0\columnwidth,angle=0]{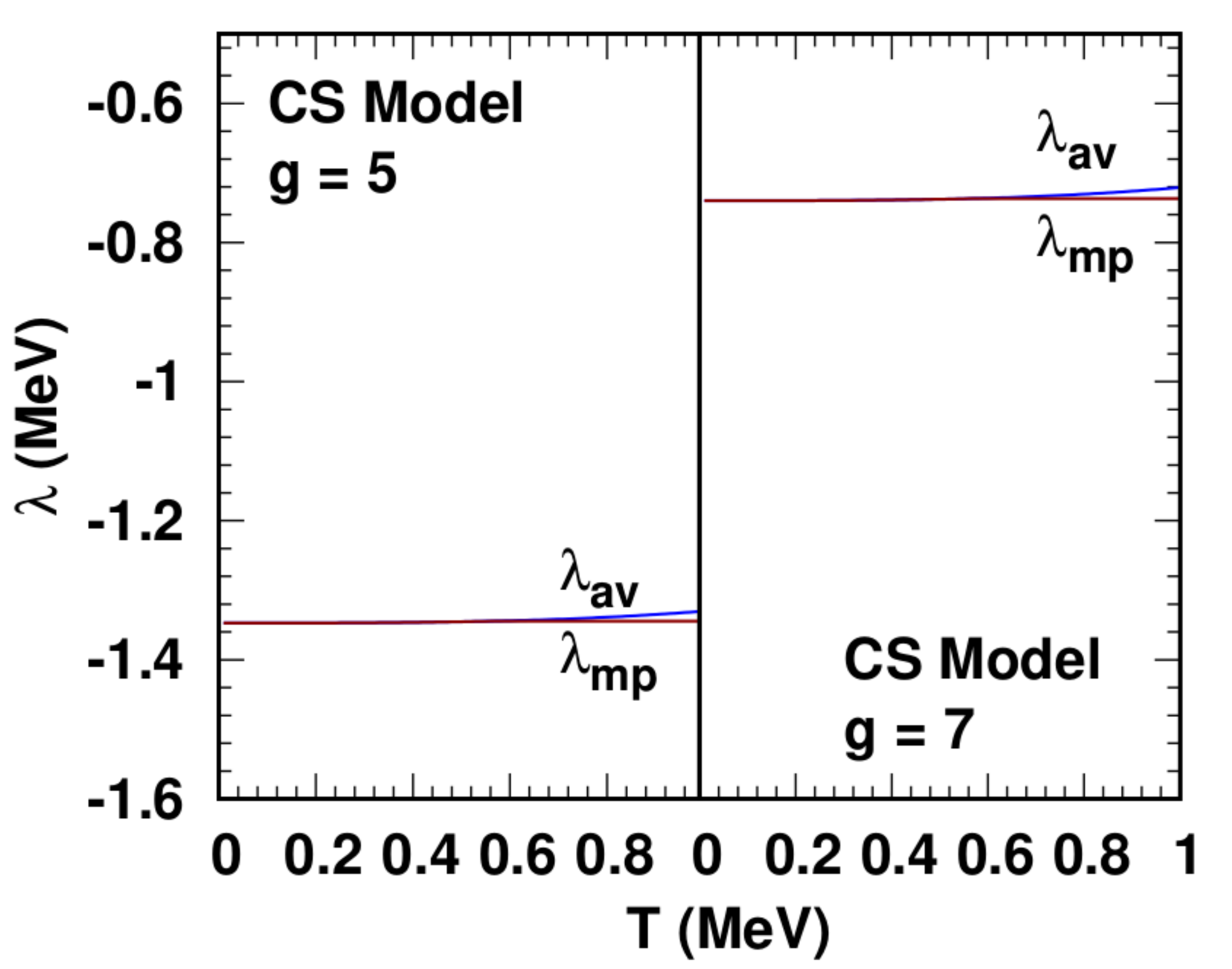}
\vspace{-0.5cm}
\caption{(Color online.) The most probable and average chemical potentials for pairing gaps, $\Delta_{\rm mp}$ and $\Delta_{\rm av}$.
\label{lambdas}}
\end{center}
\end{figure}

 Very similar results are obtained with  $g=7~{\rm MeV}^{-1}$ (as in Ref. \cite{Morettoplb}) and
$\Delta_0=1~{\rm MeV}$ for which $G=0.0462$ MeV at $T=0$ MeV.  In this case, the levels are distributed between $\pm 1.4\hbar\omega$ to ensure that roughly equal number of levels lie  above and below  $\lambda_{\rm mp}(0)=-0.7397$ MeV. For all curves shown, $\lambda(T)$ vs $T$ is calculated using Eq. (\ref{Nfluc}) prior to the calculation of $\Delta(T)$ required in the evaluation of $P(\Delta)$
in Eq. (\ref{PDelta}). The derivative $\partial\Delta/\partial\alpha$ needed in Eq. (\ref{Nfluc}) is given by \cite{Morettonucl}
\begin{eqnarray}
\frac {\partial \Delta}{\partial \alpha} &=& \frac {\sum_k (\epsilon-\lambda) (a_k-b_k) }{(\Delta/T)\sum_k (a_k-b_k)} \quad {\rm with} \nonumber \\
a_k &=& \frac 12 \frac {1}{E_k^2} \frac {1}{\cosh^2 \frac{E_k}{2T}} \quad {\rm and} \quad b_k = \frac {T}{E_k^3} \tanh \frac{E_k}{2T} \,,
\end{eqnarray}
which is valid for all models and not just for the CS model. For $T\leq T_c$, $\lambda_{\rm mp}(T)\approx \lambda_{\rm mp}(0)$ very nearly coincides with $\lambda_{\rm av}(T)$, whereas $\lambda_{\rm mp}(T)$ is slightly below $\lambda_{\rm av}(T)$ for $T > T_c$ as shown in Fig. \ref{lambdas}.

\begin{figure}[!htb]
\begin{center}
\includegraphics[width=1.0\columnwidth,angle=0]{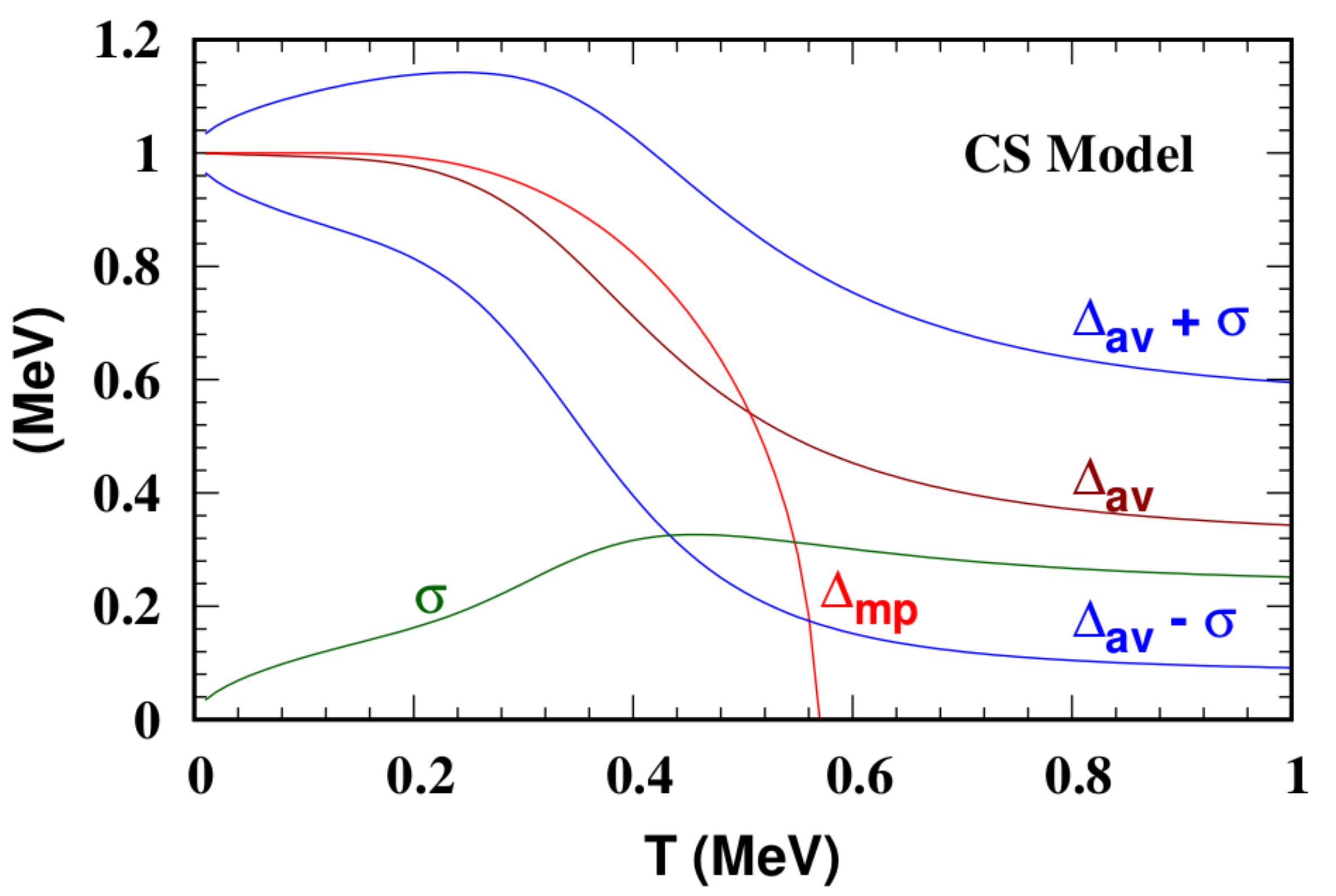}
\vspace{-0.5cm}
\caption{(Color online.) The most probable and average pairing gaps, $\Delta_{\rm mp}$ and $\Delta_{\rm av}$, along with those differing by one standard deviation, $\sigma$.
\label{CSgaps}}
\end{center}
\end{figure}

In Fig. \ref{CSgaps}, the most probable average paring gaps, $\Delta_{{\rm mp}}$ and $\Delta_{{\rm av}}$, vs $T$ are compared. Also shown are results for the standard deviation $\sigma$ and the gaps $\Delta_{{\rm av}}\pm \sigma$.   In each case, the appropriate $\lambda (T)$ was calculated with a numerical evaluation of the derivative $\partial \Delta/\partial \alpha$. 
The discontinuity at $T_c$ that occurs for $\Delta_{{\rm mp}}$ is absent for $\Delta_{{\rm av}}$ and $\Delta_{\rm {av}}\pm \sigma$. Furthermore, in the latter cases finite values of gaps persist  for $T\geq T_c$ indicating that some high-energy quasi particles continue to undergo pairing.  As first noted in Ref. \cite{Morettoplb}, these results imply that the second order phase transition present for $\Delta_{{\rm mp}}$ is considerably altered by fluctuations. 
We have verified that the qualitative features of these results are not changed when the degeneracy of each single-particle energy level is increased to 4 (for the same $\Delta_0$ and $A$).

\begin{figure}[!htb]
\begin{center}
\includegraphics[width=1.0\columnwidth,angle=0]{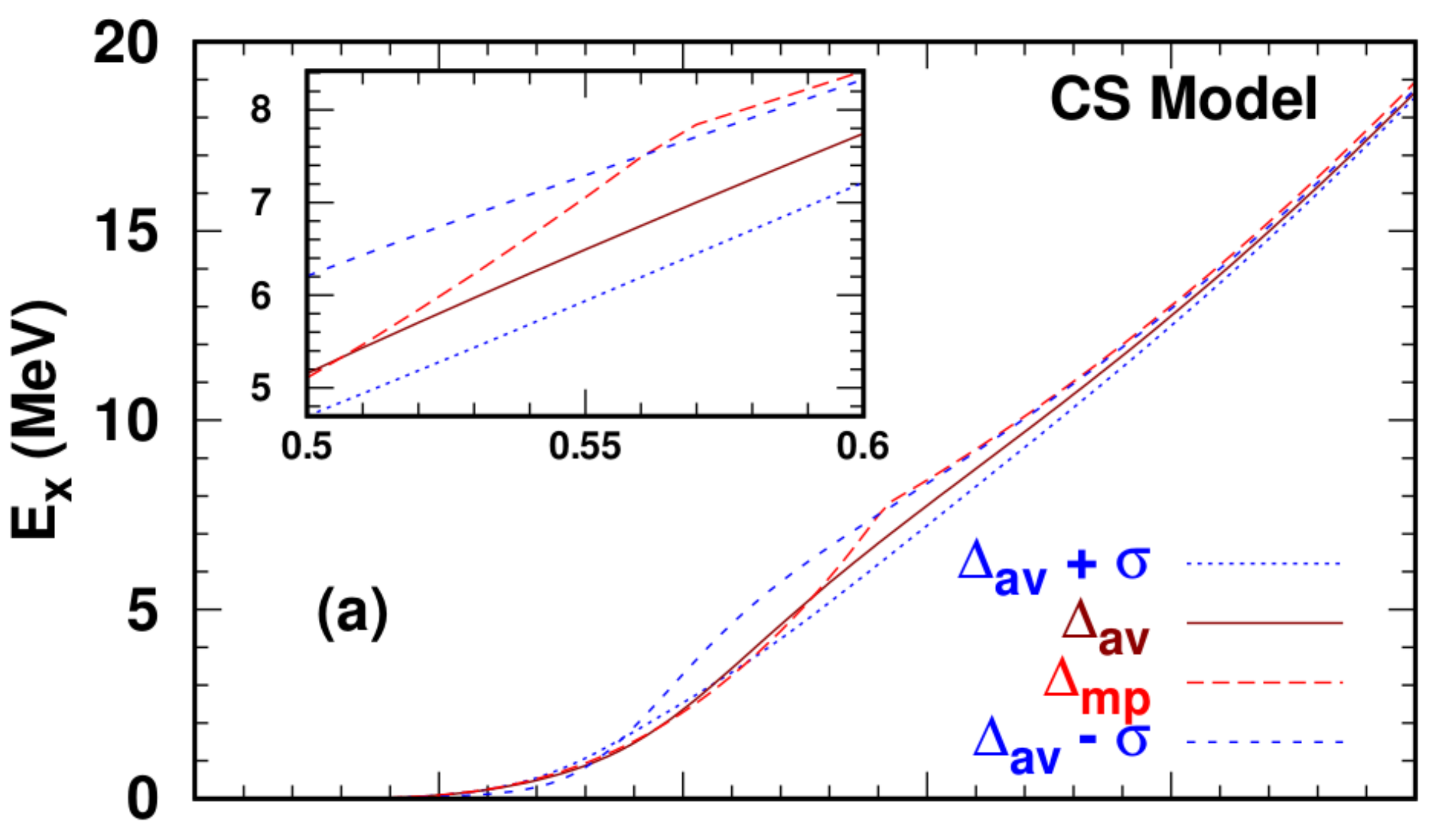}
\vspace{-0.5cm}
\label{CSCv}
\includegraphics[width=1.0\columnwidth,angle=0]{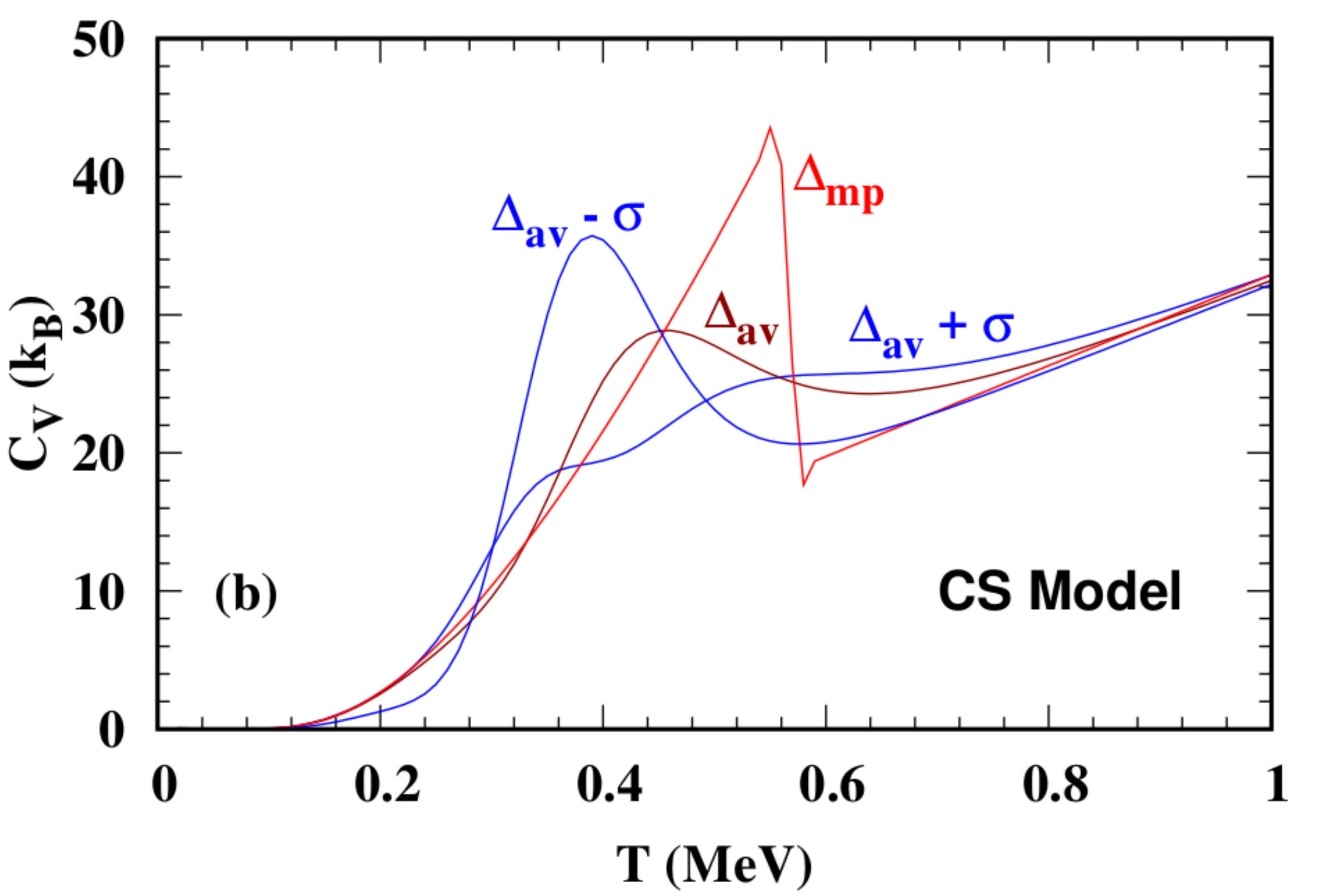}
\caption{(Color online.) (a) Excitation energies with the gaps shown in Fig. \ref{CSgaps}. (b)  Specific heats at constant volume with the gaps shown in Fig. \ref{CSgaps}.
\label{CSEx}}
\end{center}
\end{figure}

The excitation energies $E_x$ vs $T$ are shown in Fig. \ref{CSEx} for the various gap values shown in Fig. \ref{CSgaps}. The inset in Fig. \ref{CSEx} shows an expanded version of the same results in the vicinity of $T_c$. Notice that the kink present in $E_x(\Delta_{\rm mp})$ at $T_c$ is absent in all other cases as a consequence of smooth variations in $\Delta$'s around at and around $T_c$, further indicating  the lack of a strong second order phase transition. The derivative $\partial \Delta/\partial T$,  required in the evaluation $E_x=E(T)-E(0)$ from Eq. (\ref{Efluc}), is straightforwardly calculated numerically.  

The influence of fluctuations in $\Delta$ is particularly evident in the behavior of the specific heats, $C_V$'s, with respect to $T$ shown in Fig. \ref{CSEx} (b). Although the $C_V$'s with $\Delta_{\rm av}$ and $\Delta_{\rm av}\pm \sigma$ exhibit multiple extrema, the sharp discontinuity of $C_V(\Delta_{mp})$ at $T_c$ is absent.  Whether a similar behavior is exhibited in the RS and HFB models will be the subject of the next two subsections.  

\subsection*{The RS Model}

We turn now to examine the effects of fluctuations in $\Delta$ for the RS model, first with degeneracy $d=2$ and thereafter $d=2j+1$ to mimic shell-model-like configurations for $A=144$. In both cases, the uniformly distributed random sp energy levels were sorted in ascending order. The mean level spacing $\bar \delta = (\bar g)^{-1}$, where  $\bar g$ is the mean level density, was chosen to be much smaller than $\Delta_0=1$ MeV to facilitate a proper comparison with results of the CS model.  For each set of random sp energy levels,  the level separation and its probability distribution enables the calculation $\bar \delta$ and thus of  $\bar g$.

\subsection*{Degeneracy $d=2$}

The overall calculational scheme remains the same as for the CS model described above. Our results to be discussed below are for $50$ independent realizations of sp energy levels.  The probability distribution $P(\Delta)$ vs $\Delta$ in each case looks very similar to that of 
the CS Model in Fig. \ref{CSprob} and is therefore  not shown. 

\begin{figure}[!htb]
\begin{center}
\includegraphics[width=1.0\columnwidth,angle=0]{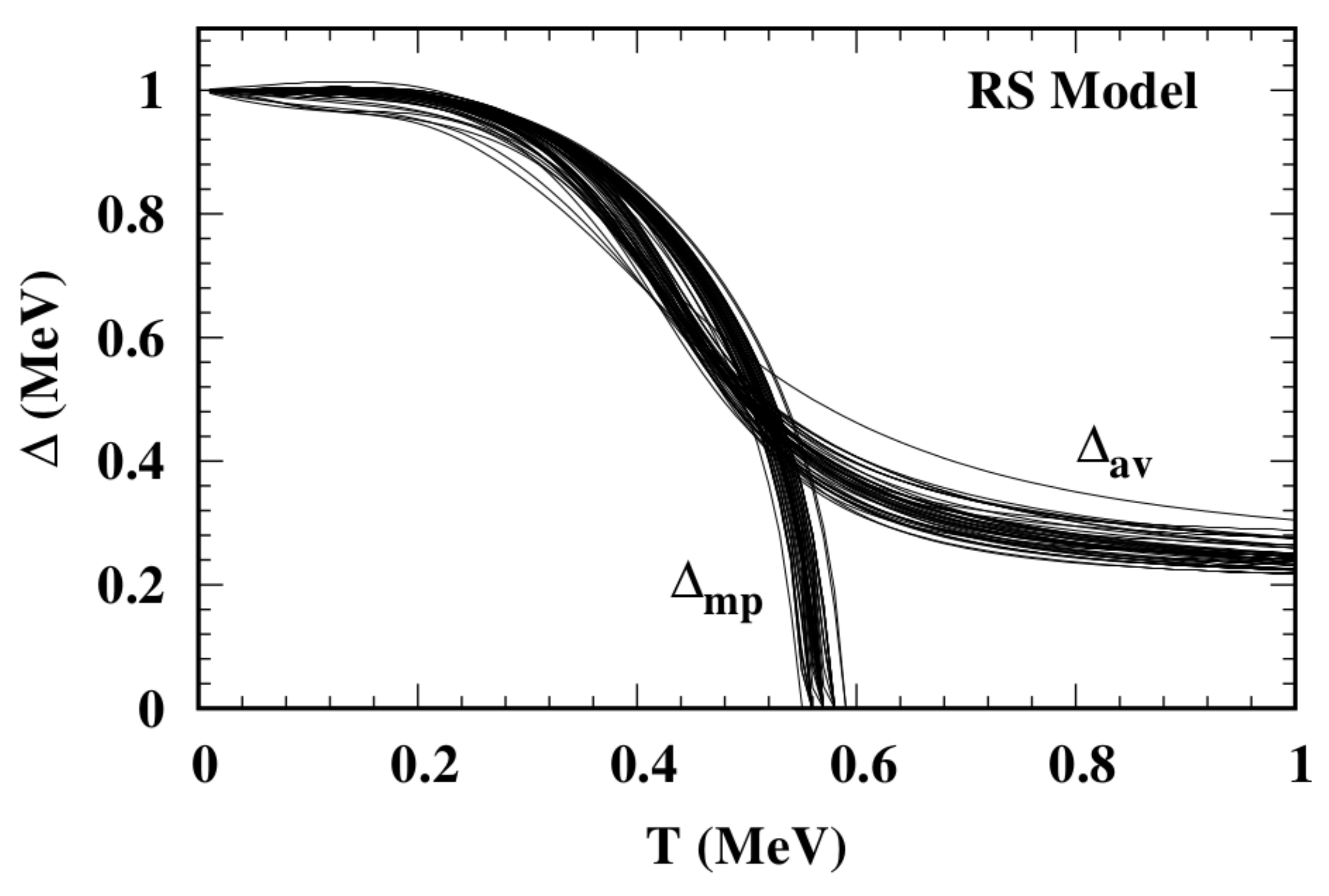}
\vspace{-0.5cm}
\caption{The most probable and average pairing gaps, $\Delta_{\rm mp}$ and $\Delta_{\rm av}$, for 50 independent random realizations of sp energies. 
\label{Gapr}}
\end{center}
\end{figure}

Figure \ref{Gapr} shows the most probable gap $\Delta_{mp}$
and the average gap $\Delta_{av}$ vs $T$ for $50$ independent realizations of sp energy levels. The standard deviations $\sigma$ 
and $\Delta_{av} \pm \sigma$ were also calculated but are omitted for visual clarity. The band structures for $\Delta_{\rm mp}$  and  $\Delta_{\rm av}$ establish statistical bounds for each quantity.  As for the CS model, $\Delta_{\rm av}$  lacks the sharp discontinuity at $T_c$ and persists with a non-vanishing gap above $T_c$.

\begin{figure}[!htb]
\begin{center}
\includegraphics[width=1.0\columnwidth,angle=0]{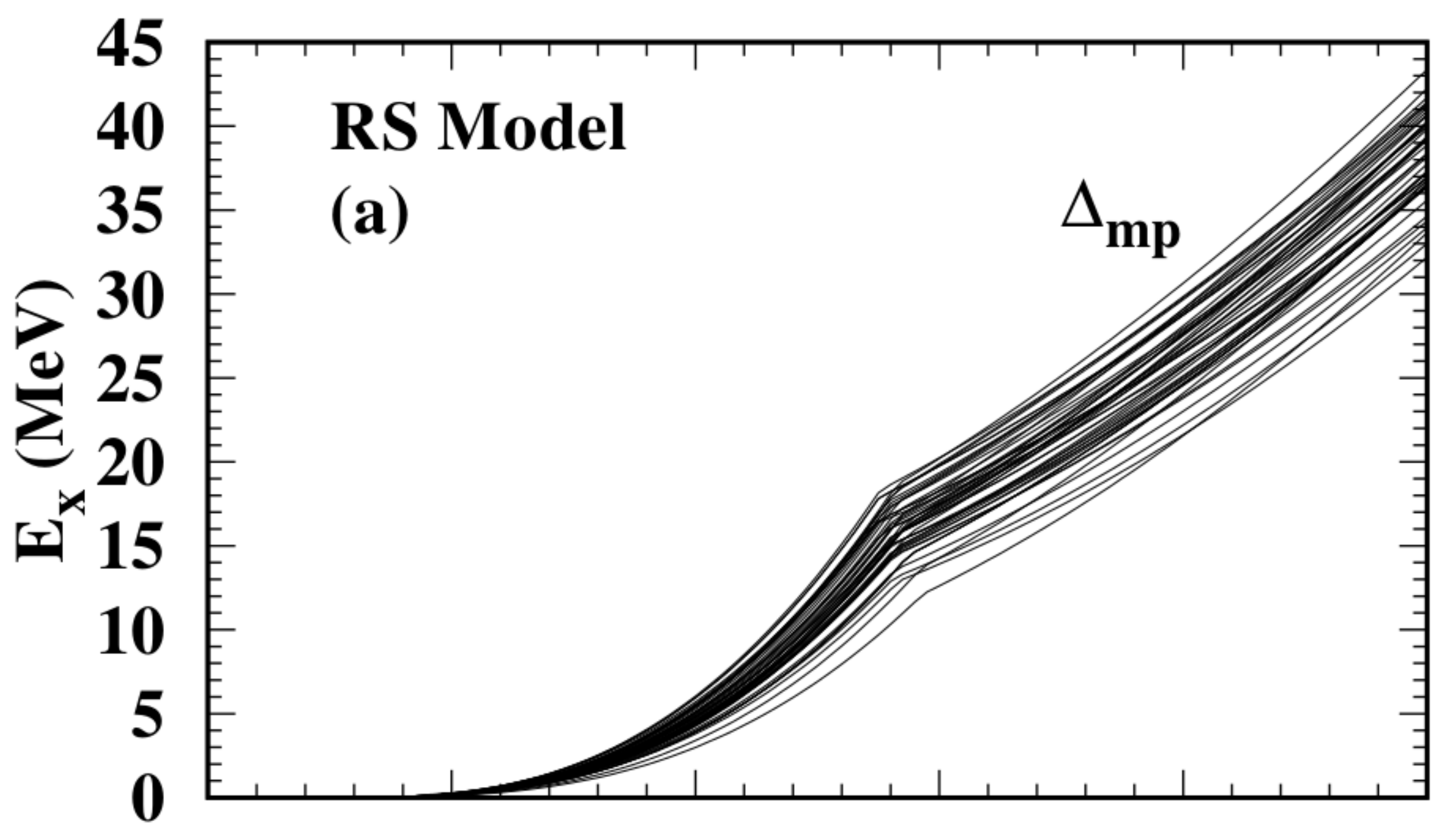}
\vspace{-0.50cm}
\includegraphics[width=1.0\columnwidth,angle=0]{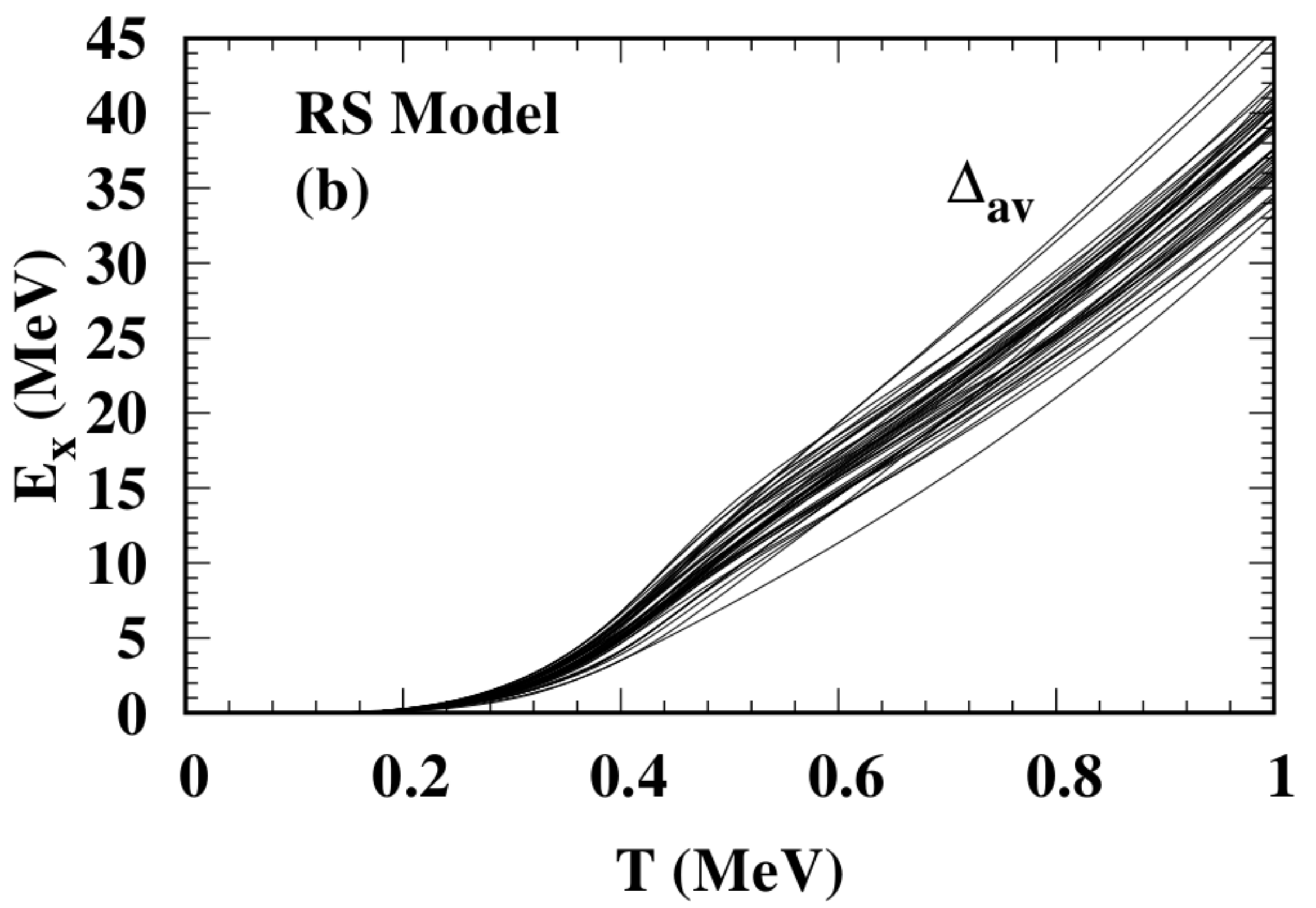}
\end{center}
\caption{Excitation energies with (a) $\Delta_{\rm mp}$ and (b) $\Delta_{\rm av}$ shown in Fig. \ref{Gapr}.
\label{Exr}}
\end{figure}

The excitation energies calculated using  $\Delta_{mp}$ and $\Delta_{av}$ from Eq. (\ref{Efluc}) are shown in Figs. \ref{Exr}(a) and (b), respectively.  
Particle number
was conserved at every stage of the calculation by using the extended number equation Eq. (\ref{Nfluc}). The kinks in  
$E_x(\Delta_{\rm mp})$ at $T_c$ are absent in $E_x(\Delta_{\rm av})$ (Figs. \ref{Exr} (a) and (b) ),  again signifying the lack of a second order phase transition.  
  
\begin{figure}[!htb]
\begin{center}
\includegraphics[width=1.0\columnwidth,angle=0]{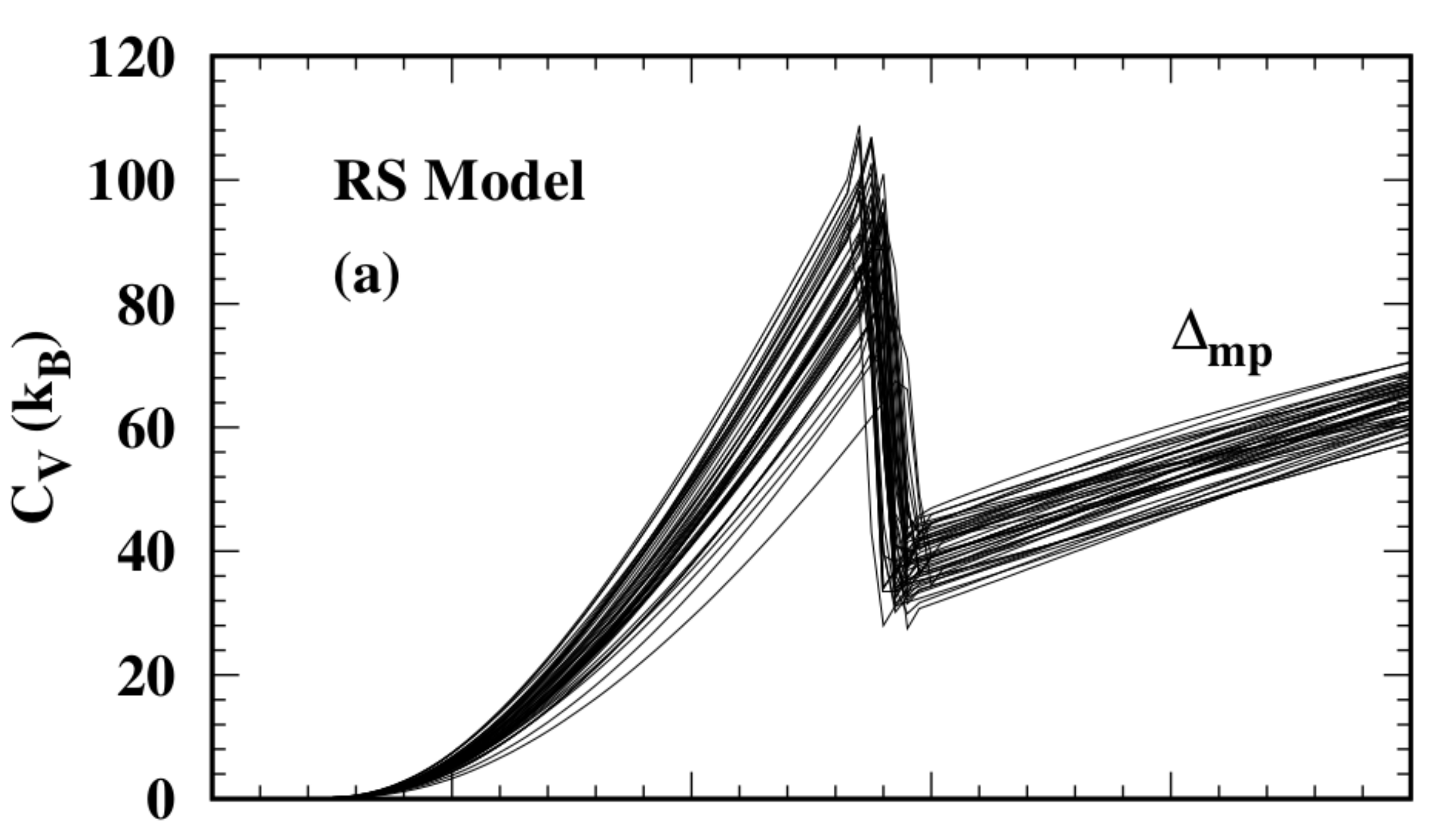}
\vspace{-0.40cm}
\includegraphics[width=1.0\columnwidth,angle=0]{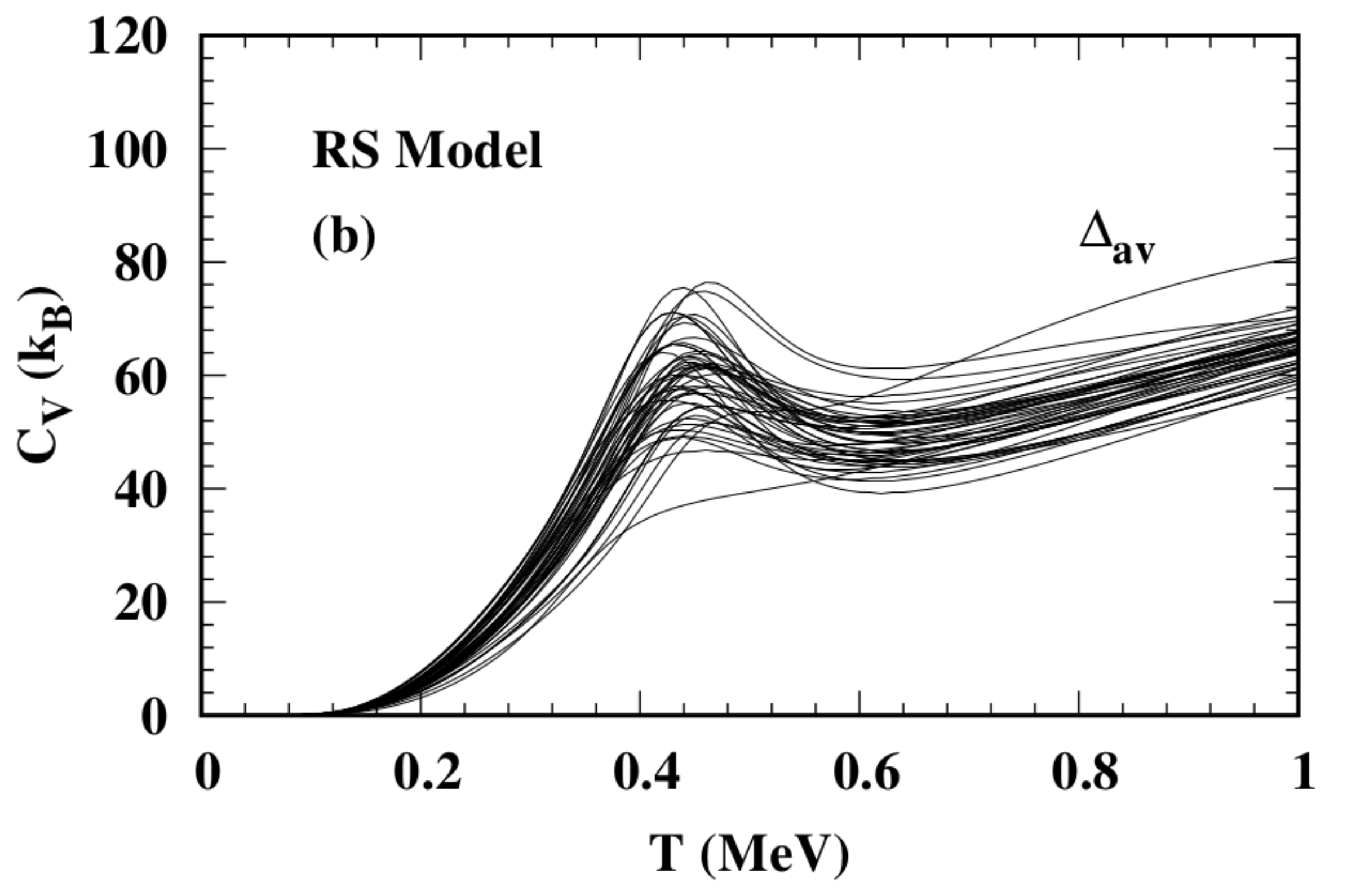}
\caption{Specific heats at constant volume with (a) $\Delta_{\rm mp}$ and (b) $\Delta_{\rm av}$ shown in Fig. \ref{Gapr}.
\label{Cvr}}
\end{center}
\end{figure}

The corresponding specific heats, calculated by taking numerical derivatives, are shown in Figs. \ref{Cvr}(a) and (b). The discontinuity
in $C_V$ present for all different sets sp energy levels when using $\Delta_{\rm mp}$ is absent when $\Delta_{\rm av}$, likely more appropriate for systems with small number of particles for which fluctuations are large, is used.
The discontinuity  is replaced by a 
so-called ``shoulder-like'' structure, which points to the persistence of pairing correlations but not a second order phase transition. Note that the qualitative features for all thermodynamic quantities in the RS model with $d=2$ are similar to those of the CS Model. 

\subsection*{Degeneracy $d=2j+1$ }

Inclusion of angular momentum degeneracy $d=2j+1$ in the sp levels of the RS model makes  the model  to better mimic nuclei. In what follows, 36 sp energy levels were generated between $\pm 2\hbar\omega$ using a uniform sequence random number generator and then sorted in ascending order so that the lowest energy level is at the bottom. The sorted energy levels were then assigned individual shell model-like degeneracies  2, 4, 6, etc. For each set of  a large number of such realizations, the number and gap equations were then solved for $T=0$ and $\Delta_0 = 1$ MeV for a fixed $N$ using the sp energy levels to extract the corresponding pairing strength $G$ and Fermi energy $\lambda_0$. To ensure pairing as a Fermi surface phenomenon, approximately equal number of energy levels are needed above and below the Fermi energy. Consequently, all the energy levels were then shifted by a constant energy so that the shifted Fermi energy $\lambda_s$ is slightly below 0 MeV as a hole state.

\begin{figure}[!htb]
\begin{center}
\includegraphics[width=1.0\columnwidth,angle=0]{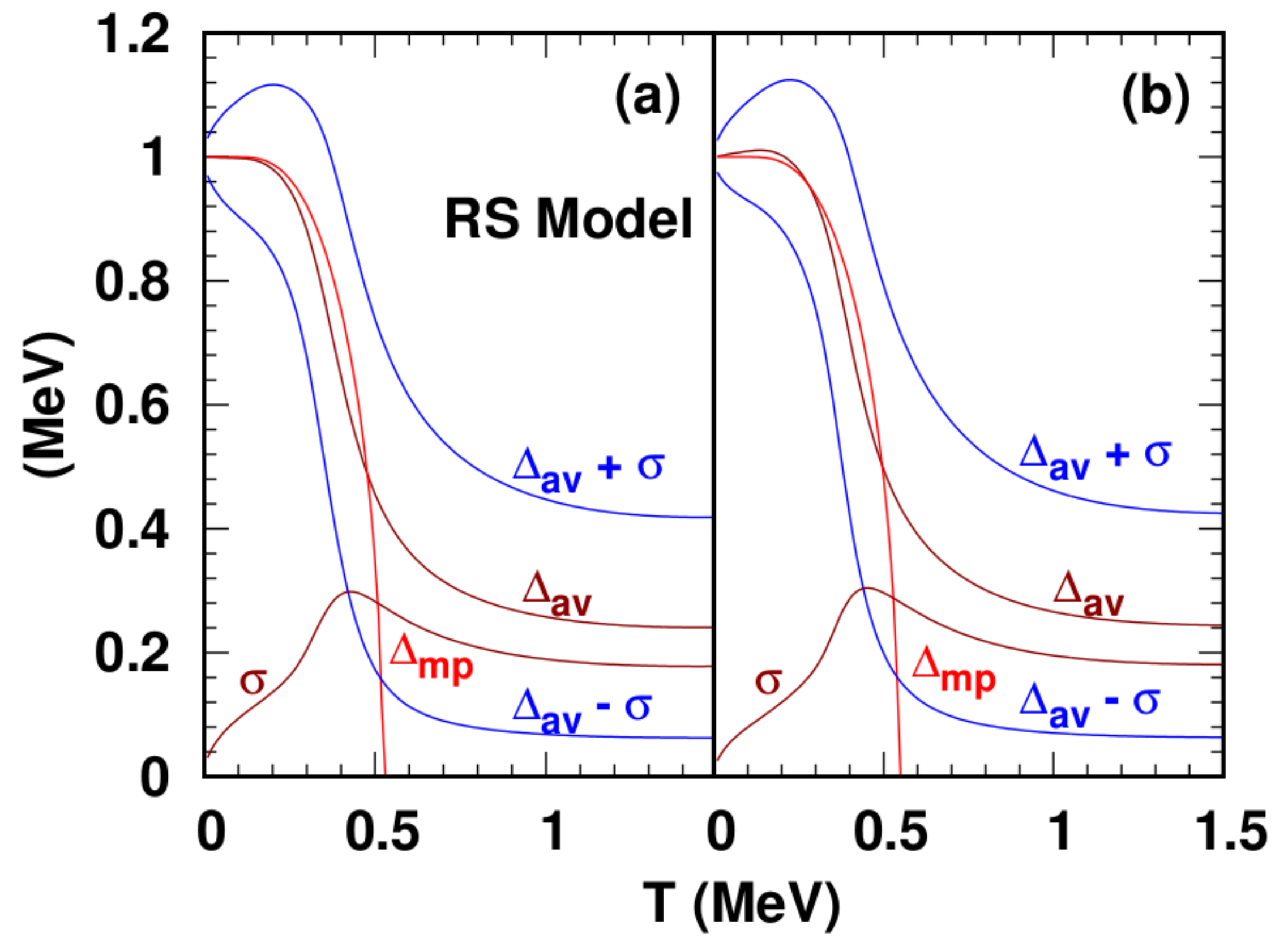}
\vspace{-0.5cm}
\caption{(Color online.) Same as Fig. \ref{Gapr}, but for the RS model with degeneracy $d=2j+1$. 
\label{RSgap}}
\end{center}
\end{figure}

The results of  $\Delta_{\rm mp}, \Delta_{\rm av}$ and $\Delta_{\rm}\pm \sigma$ for two such calculations as described above among hundreds of individual random realizations of sp energy levels 
are shown in Fig. \ref{RSgap}. 
The latter two quantities were calculated following  the procedure described at the beginning of this section.   
For the cases shown,  the average level spacing $\bar\delta$ was found to be $0.82$ and $0.87$ MeV,  
respectively, which are slightly less than the zero temperature gaps $\Delta_0 = 1$ MeV.   These numbers make a semiclassical treatment of fluctuations valid, albeit on the borderline of requiring a quantum treatment needed for cases in which $\bar\delta \gtrsim \Delta_0$ as found for many nuclei.  Note that the qualitative features in Fig. \ref{RSgap} are 
similar to those in Fig. \ref{CSgaps} of the CS Model. 

\begin{figure}[!htb]
\begin{center}
\includegraphics[width=1.0\columnwidth,angle=0]{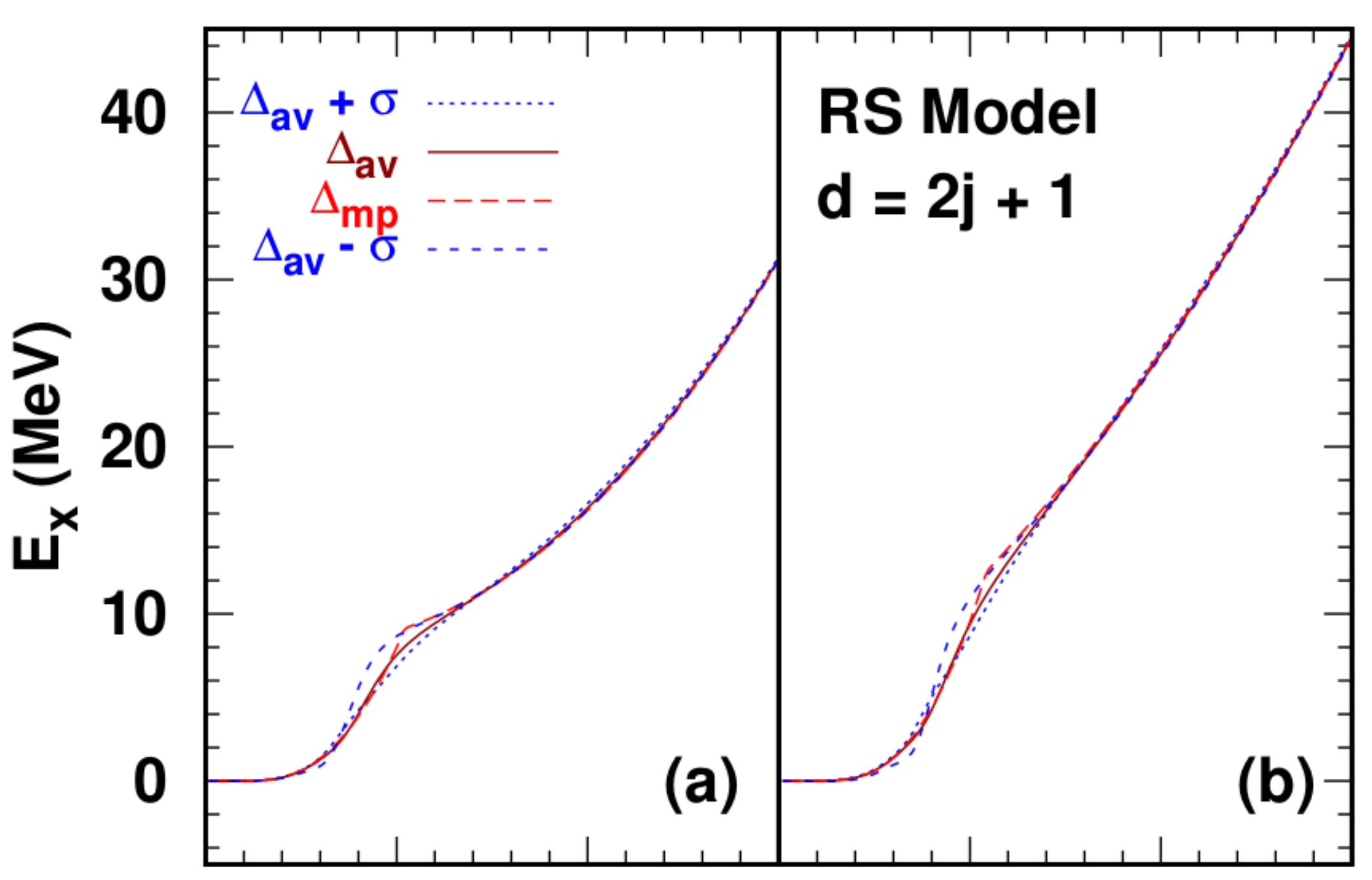}
\vspace{-0.50cm}
\includegraphics[width=1.0\columnwidth,angle=0]{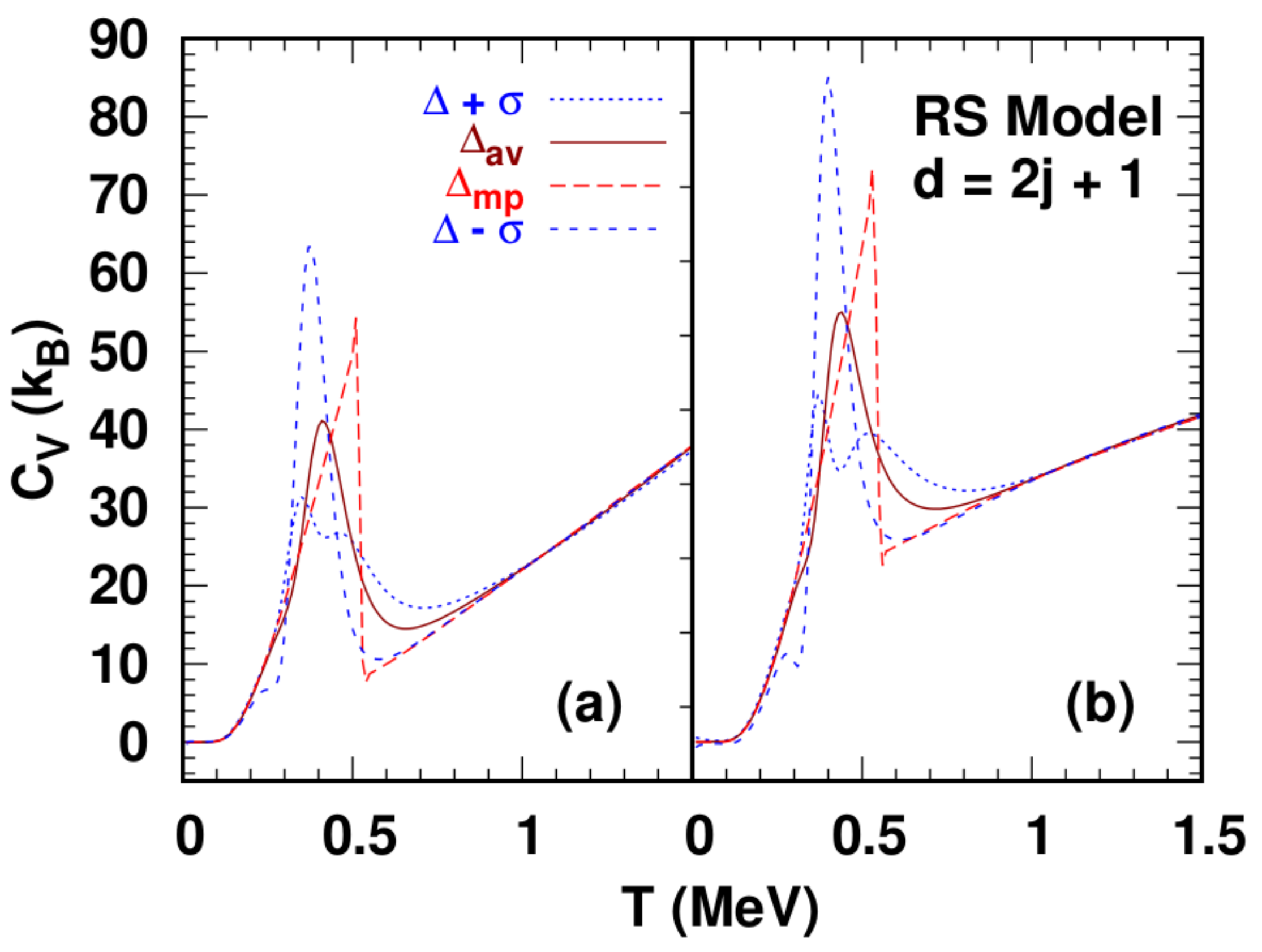}
\vspace{0.0cm}
\caption{(Color online.) Same as Fig. \ref{CSEx}, but for the RS model with degeneracy $d=2j+1$. } 

\label{RSExcv}
\end{center}
\end{figure}

The excitation energies and specific heats corresponding to the results in Fig. \ref{RSgap} are shown in Fig. \ref{RSExcv}(a)-(d).  Although the overall features in this figure seem very similar to those of the CS Model, values of $E_x$ and its slope with respect to $T$ ($C_V)$ are different owing to the different bunching and degeneracy of the individual sp energy levels of the RS model. 
As the number of particles in the two cases are fixed at $N=144$,  differences between the two cases reflect the different dispositions of the sp energy levels which can arise due to use of different energy density functionals in describing the same nucleus. One noticeable feature is that the $E_x$ curve calculated using 
$\Delta_{\rm av}$-$\sigma$ obtains slightly negative values for near zero temperatures. There is a 
possibility of  similar occurrence even for $E_x(\Delta_{av})$ for other sp energy realizations. This behavior can be attributed to the failure of a semiclassical treatment in the very low temperature region.

The specific heat curves in Figs. \ref{RSExcv}(c)-(d)  again show the smoothing 
effect of fluctuations. The ``shoulder-like'' structures evident when fluctuations are incorporated 
as opposed to the sharp discontinuity in $C_V(\Delta_{mp})$ indicate the absence of a second order phase transition. This is a very close representation of the situation in nuclei as found in experiments. 

\subsection*{HF calculations for Nuclei}

\begin{figure}[!htb]
\begin{center}
\includegraphics[width=1.0\columnwidth,angle=0]{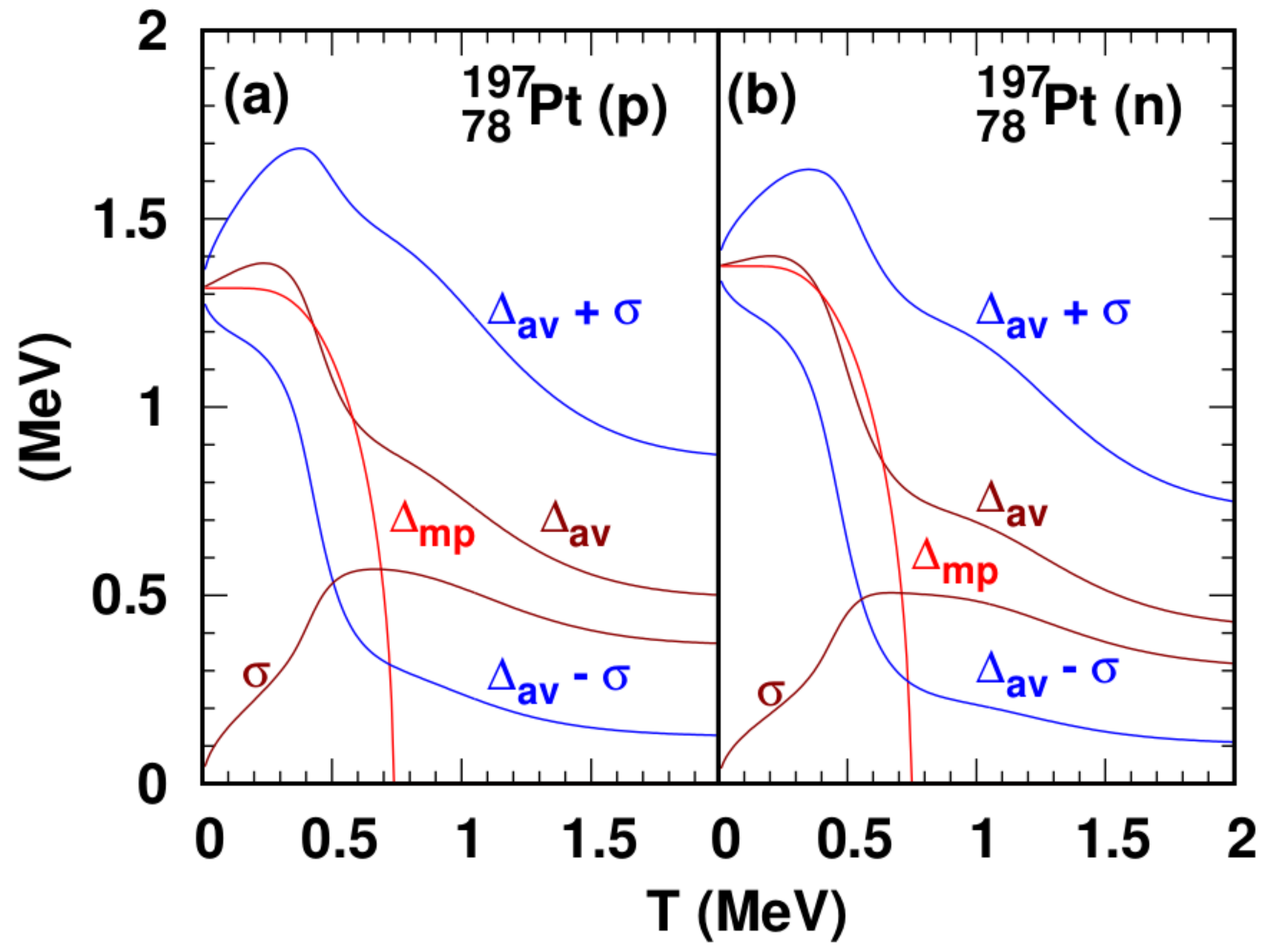}
\vspace{-0.5cm}
\caption{(Color online.) Same as Fig. \ref{CSgaps}, but for protons (a) and neutrons (b) in $^{197}$Pt. 
\label{Gapnp197}}
\end{center}
\end{figure}

In this section, results of HF calculations for the odd-even nucleus $\rm{^{197}_{78}Pt}$ are compared with those of the CS and  RS models. 
Pairing properties were calculated within the BCS formalism with a constant force for illustrative purposes. 
Neutrons and protons were treated as two separate systems, but owing to the linearity of thermodynamic quantities
they can be simply added to obtain the same thermodynamic quantities for the whole nucleus. Figure
\ref{Gapnp197} shows the proton and neutron gaps vs $T$. The most probable gaps $\Delta_{\rm mp}$  for protons and neutrons at $T=0$  were calculated by fixing the coupling strengths $G$ so that the gaps conform to the systematics indicated by Eqs. (\ref{nsystematics}) and (\ref{psystematics}). Also shown are $\Delta_{\rm av}$ 
along with their standard deviations as a function of temperature.
These results have qualitative resemblance with those of the CS and RS models.

\begin{figure}[!htb]
\begin{center}
\includegraphics[width=1.0\columnwidth,angle=0]{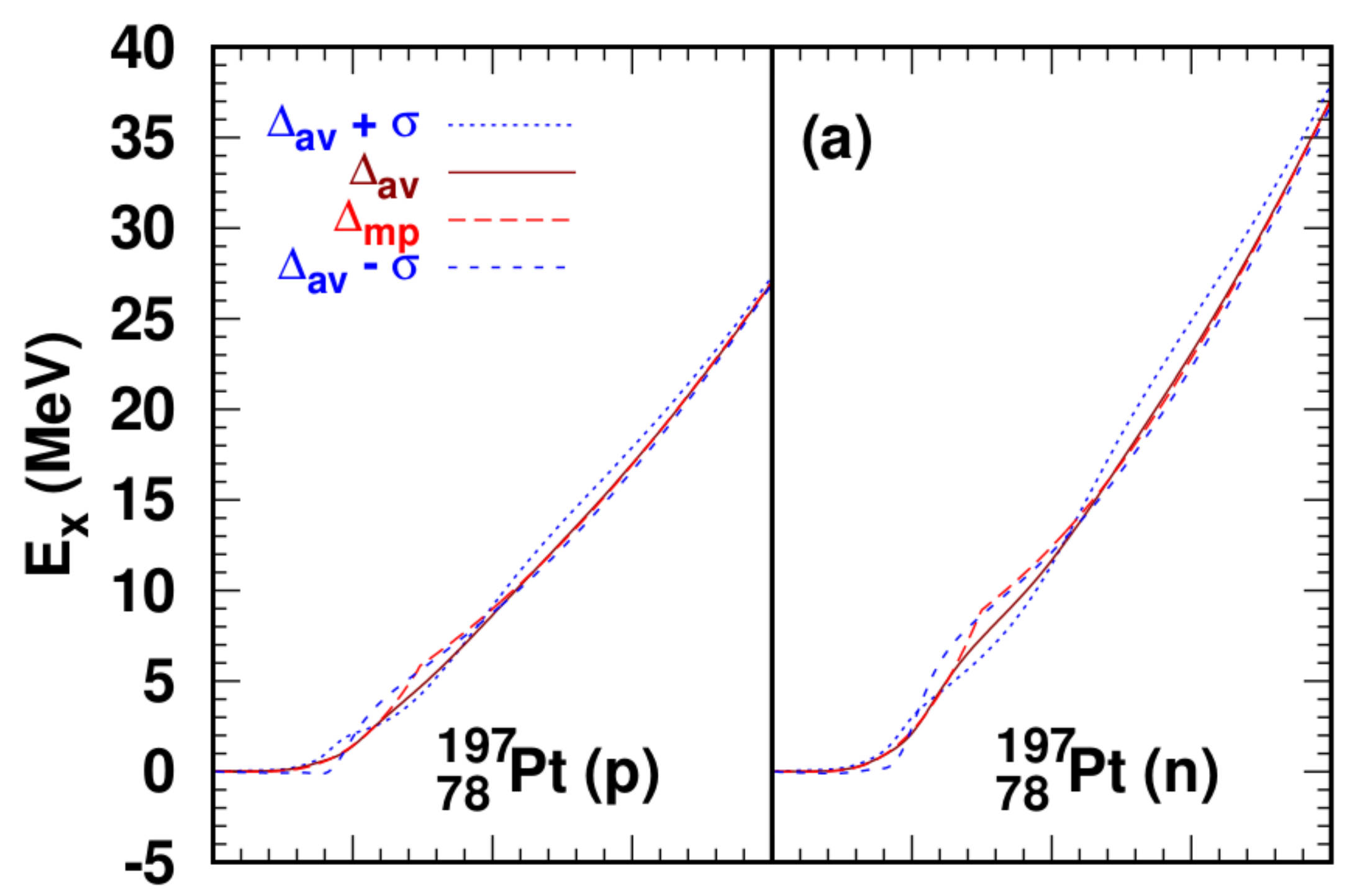}
\vspace{-0.50cm}
\includegraphics[width=1.0\columnwidth,angle=0]{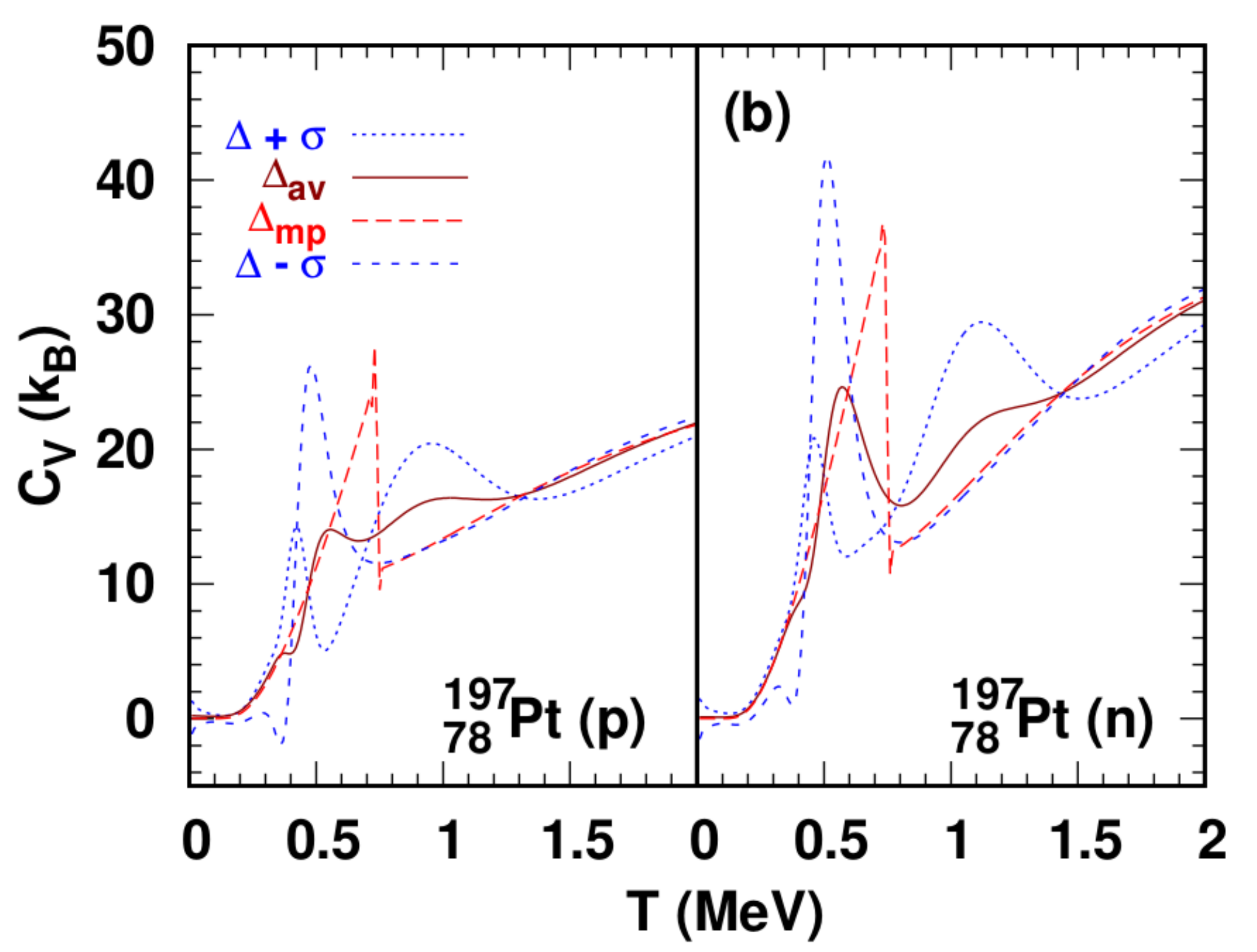}
\end{center}
\caption{(Color online.)  Same as Fig. \ref{CSEx}, but for protons and neutrons in $^{197}$Pt. 
\label{Cvrnp197}}
\end{figure}

The excitation energy and specific heat curves shown in Fig. \ref{Cvrnp197} also show similar qualitative
behavior to those of the RS Model. A noticeable feature is the  larger 
fluctuations than for the RS Model. This is owing to $\bar\delta$ being $1.97$ \& $1.68$ MeV, respectively, for 
protons and neutrons. These values of $\bar\delta$ are larger than the corresponding $\Delta_0$'s which indicate that improvement over the mean field BCS  
treatment, which advocates use of most probable gaps, is necessary \cite{Anderson59,Falci00,LLI,Morettoplb,Al13}.   
Results of our semiclassical treatment of fluctuations in the RS Model as well those in the HF+BCS calculations with a constant force  
highlights that pairing correlations persist even if a second order phase transition disappears. 
A similar semiclassical treatment of pairing correlations with similar results for $^{94}{\rm Mo}$ using Nilson model sp energy levels can be found in Ref. \cite{Kargar13}.  
Analogous results have been obtained with  more advanced treatments that include improvements such as HFB calculations beyond mean field theory and a quantum  treatment of fluctuations (see below and the many articles in Ref. \cite{Fiftyyrs}).

Detailed comparisons with experiments are premature at this development stage of the RS model. The influence of additional sources of fluctuations such as beyond mean field effects, collective effects and those from rotation should be considered in a fully quantum treatment  to provide a comparison with the semiclassical treatment adopted in this work. We expect further modifications of the shoulder-like or the S-Shaped structure when these and additional sources of fluctuations are included.

\subsection*{Beyond mean field theory}

The HF theory includes pairing in nuclei using the BCS approximation by treating the pair correlations 
through time-reversed orbital wave functions. In this approach, the HF equations are self-consistently solved
to find variational minima using an underlying energy density functional. But the minima (HF wave functions) so obtained using HF/BCS 
could be different from that of HFB \cite{RB13}. This is due to the more complete wave functions of the Bogoliubov 
transformation in contrast to the small configuration space HF wave functions. Hence, HFB is more inclusive of
physical effects than HF.

In the HFB/BCS approach, broken symmetries which are artifacts of  the mean field approximation appear. Beyond mean field techniques restore the number symmetry  and treat fluctuations in the BCS order 
parameter; see Ref. \cite{RB13} and references therein. Other popular techniques include the random phase 
approximation (RPA) and their derivatives \cite{Shimizu00}. Even then, 
many technical difficulties, such as the sign of the overlap of HFB wave functions and additional 
difficulties with odd-A nuclei, arise \cite{RB13}. 
Correlations beyond the mean field have also been treated in Ref. \cite{Al13} by using the  Hubbard-Stratonovich  (HS) transformation which  can be incorporated 
in many different ways, e.g.,  the static-path approximation (SPA) in which only the thermal fluctuations are 
addressed. The SPA coupled with RPA includes time-dependent quantal fluctuations in addition to thermal effects. 
Advanced, but computationally intensive methods such as the Auxiliary-Field Monte Carlo (AFMC) approach include additional
fluctuations \cite{Al13}.  This reference gives an account of the various methods employed to treat fluctuations at the quantum level.

While the static (BCS or HFB) mean field approximation is an adequate  
treatment for a  spherical or non-rotating nucleus, dynamic effects (such as pairing 
vibrations) need to be included on top of the static  mean field for a rotating nucleus.
The effect of pairing on 
rapidly rotating nuclei is to significantly reduce the rigid body moment of inertia. The formation of 
Cooper pairs means having two nucleons with time-reversed conjugate orbits. In rapidly rotating nuclei, 
nucleons are forced to align their angular momenta with the rotation axis which leads to the breaking of 
Cooper pairs. This results in a gradual decrease of the effective pairing gap (static gap + dynamic gap) 
as opposed to a sharp disappearance of the static gap, see, e.g., \cite{Shimizu00}. 

In this work, we have examined the role of thermal fluctuations in the RS model using a semiclassical treatment. 
A quantitative comparison with a quantum treatment that includes additional sources of fluctuations within the RS model is beyond the scope of this paper, but will be reported in a separate work.

\section{Summary and Conclusions}
\label{Sumconc}

We turn now to summary and conclusions. 
In the medium-to-heavy mass region, spherical and deformed nuclei accessible to laboratory experiments, and particularly those
only realized in the highly neutron-rich environments encountered in astrophysical phenomena, are characterized by an
assortment of bunched single-particle (sp) energy levels owing to shell and pairing effects. Laboratory experiments performed on various nuclei have revealed a shoulder-like structure around the critical temperature $T_c$ expected from a second order phase transition from the BCS formalism of the pairing phenomenon  involving fermions, but not a discontinuous jump in the specific heat from the paired to the normal phase \cite{quenching,sn116,toft,pygmy}.

The main contribution of this work is the introduction of the random spacing (RS) model in which the sp energy levels are distributed around the Fermi energy to mimic those of nuclei obtained via the use of different energy density functionals. 
The distributions of these sp energy levels closely resemble those of randomly generated levels around the Fermi surface. 
Exploiting this similarity, we have calculated the basic characteristics of the pairing correlations in the RS model and compared the results with those of select nuclei.  Aspects of the RS model are studied in two distinct stages as summarized below.

In the first stage,  the BCS formalism,  which employs the most probable pairing gaps to calculate the critical temperature,  the behaviors of the entropy and specific
heat at constant volume as functions of temperature (excitation energy) and angular momentum, is  used for the  sp  
energy levels of the RS model. Comparisons with results of the Fermi gas, constant spacing models and nuclei are provided. 
Our principal results at this stage are as follows. From the statistically-based bounds obtained, we find that  the ratio of the critical temperature to the zero-temperature pairing gap is close to its  Fermi gas value, and appears to be a robust result. However,  the ratio of the paired to normal phase specific heats at the critical temperature $T_c$ differs significantly from its Fermi gas counterpart. The scatter around the mean value for the discontinuity in the specific heat at the critical temperature is largest when a couple of sp levels lie closely on either side of the Fermi surface, but other levels are far away from it.

In the second stage, the role of fluctuations, expected to be large for systems with small number of particles, is studied.  Based on a semiclassical treatment of thermal fluctuations first developed in Ref. \cite{Morettoplb} for the CS model and later applied with some improvements in Ref. \cite{Kargar13} for $^{94}{\rm Mo}$,   applications are considered here to the RS model.   The chief result of this investigation is that the second order phase transition, a consequence of using  the most probable values for the paring gaps in the BCS formalism, is suppressed and replaced by a shoulder-like structure around $T_c$ when the average values for the pairing gaps are used indicating the lasting presence of pairing correlations.  Such a structure is indeed observed in experiments performed on several nuclei \cite{quenching,sn116,toft,pygmy}.  We note, however, that a semiclassical treatment is strictly valid only when the mean sp level spacing around the Fermi surface is smaller or nearly equal to the zero temperature pairing gap  and a fully quantum treatment of fluctuations becomes necessary otherwise to overcome the limitations of the BCS formalism  \cite{Anderson59,Falci00,LLI,Morettoplb,Al13}.  Contrasting the semiclassical and quantum treatments of fluctuations  as well as 
investigations of fluctuations in highly neutron-rich isotopes with more advanced techniques 
in the context of the RS model will be undertaken in  future works. 

To the extent that the sp levels of the RS model resemble those of nuclei that exhibit considerable dependence on choices of the energy density functionals and pairing schemes used, our results indicate the variation to be expected in the basic characteristics of the pairing phenomenon in nuclei.  These results can help to perform sensitivity tests in astrophysical settings which harbor exotic nuclei.

\section*{acknowledgments}
We thank P.-G. Reinhardt for providing us with computer programs to generate single-particle energy levels for nuclei based on  various pairing schemes. Beneficial conversations with Steve Grimes, Alexander Voinov and Tom Massey are gratefully acknowledged. This work was performed with research support from the U.S. DOE grant No. DE-FG02-93ER-40756.

\bibliographystyle{h-physrev3}
\bibliography{PRLreferences}

\end{document}